\documentclass[a4paper]{JHEP3}

\usepackage{amsmath,feynmp,slashed,subfig}
\title{Z boson decay to photon plus Kaluza-Klein graviton in large extra
  dimensions}

\author{Benjamin C.~Allanach and Jordan P.~Skittrall \\ Department of
  Applied Mathematics and Theoretical Physics, \\ Centre
  for Mathematical Sciences, Univeristy of Cambridge, \\ Wilberforce Road, Cambridge, CB3~0WA,
  United Kingdom\\E-mail: \email{B.C.Allanach@damtp.cam.ac.uk},
  \email{J.P.Skittrall@damtp.cam.ac.uk}}

\author{K.~Sridhar \\ Department of Theoretical Physics, \\ Tata
  Institute of Fundamental Research, \\ Homi Bhabha Road, Bombay
  400005, India\\E-mail: \email{sridhar@theory.tifr.res.in}}

\abstract{In the large extra dimensional ADD scenario, Z~bosons
  undergo a one-loop decay into a photon and Kaluza-Klein towers of
  gravitons/gravi-scalars. We calculate such a decay width, extending
  previous arguments about the general form of the four-dimensional
  on-shell amplitude. The amplitudes calculated are relevant to
  processes in other extra dimensional models where the Standard Model
  fields are confined to a 4-brane.}


\preprint{DAMTP-2007-42 \\ TIFR/TH/07-08}

\newcommand{\htil}{\tilde{h}_{\lambda \rho}^{\vec{n}}}
\newcommand{\htilup}{\tilde{h}^{\lambda \rho,\vec{n}}}

\newcommand{\htiln}{\tilde{h}^{\vec{n}}}

\newcommand{\ptil}{\tilde{\phi}_{ij}^{\vec{n}}}

\newcommand{\ptiln}{\tilde{\phi}^{\vec{n}}}

\newcommand{\matel}{\mathcal{M}}

\newcommand{\polg}{\mathcal{E}^{\lambda\rho}(q)}

\newcommand{\polgr}{\mathcal{E}^{\rho\lambda}(q)}

\newcommand{\polgs}{\mathcal{E}^{\lambda\rho *}(q)}

\newcommand{\polp}{\varepsilon^{\nu}(k)}

\newcommand{\polps}{\varepsilon^{\nu *}(k)}

\newcommand{\polz}{\varepsilon^{\mu}_{\rm{Z}}(p)}

\newcommand{\kkm}{m_{\vec{n}}}

\newcommand{\allconthf}{T_{\lambda\rho\mu\nu}^{(hf)}}
\newcommand{\diagconthf}[1]{T_{\lambda\rho\mu\nu}^{(hf:{\rm #1})}}
\newcommand{\allconthw}{T_{\lambda\rho\mu\nu}^{(hW)}}
\newcommand{\diagconthw}[1]{T_{\lambda\rho\mu\nu}^{(hW:{\rm #1})}}
\newcommand{\allconthc}{F_{\lambda\rho\mu\nu}^{(h\times)}}
\newcommand{\diagconthc}[1]{F_{\lambda\rho\mu\nu}^{(h\times :{\rm #1})}}

\newcommand{\allcontpf}{T_{\mu\nu}^{(\phi f)}}
\newcommand{\diagcontpf}[1]{T_{\mu\nu}^{(\phi f:{\rm #1})}}
\newcommand{\allcontpw}{T_{\mu\nu}^{(\phi W)}}
\newcommand{\diagcontpw}[1]{T_{\mu\nu}^{(\phi W:{\rm #1})}}
\newcommand{\allcontpc}{F_{\mu\nu}^{(\phi \times )}}
\newcommand{\diagcontpc}[1]{F_{\mu\nu}^{(\phi \times :{\rm #1})}}

\newcommand{\lint}{\int \frac{d^4 l}{(2\pi)^4}}
\newcommand{\linttr}{\lint \hbox{Tr} \left[ \tilde{\gamma}_{\mu}}
\newcommand{\ldint}{\int \frac{d^D l}{(2\pi)^D}}

\newcommand{\fmfframevertex}[1]{\fmfframe(0,0.7)(0.25,0.5){#1}}

\newcommand{\fmfplab}[2]{\fmfv{label=$ #1 $,l.dist=0.6mm}{#2}}

\newcommand{\twozs}{\left( Z_{ZZ}^{1/2}Z_{ZA}^{1/2} \right)}
\newcommand{\fourzs}{\left( Z_{ZZ}^{1/2}Z_{ZA}^{1/2}+Z_{AZ}^{1/2}Z_{AA}^{1/2}\right)}
\newcommand{\msdms}{\left( M_Z^2 + \delta M_Z^2 \right)}
\newcommand{\twozsnobracket}{Z_{ZZ}^{1/2}Z_{ZA}^{1/2}}
\newcommand{\fourzsnobracket}{Z_{ZZ}^{1/2}Z_{ZA}^{1/2}+Z_{AZ}^{1/2}Z_{AA}^{1/2}}

\newcommand{\md}{\mu^{(4-D)}}

\newcommand{\dimregid}[3]{\md \ldint \frac{#1}{(l^2-X^2)^3} &= #2
    \frac{i}{(4\pi)^2} \left[ \frac{2}{\epsilon} -
    \gamma + \log (4\pi) - \log \left( \frac{X^2}{\mu^2} \right) #3
    \right]}

\newcommand{\polsumgel}[2]{\left( \eta^{#1 #2} -
  \frac{q^{#1}q^{#2}}{\kkm^2} \right)}

\newcommand{\thvloop}[4]{%
\begin{fmfgraph}(3,2)
\fmfpen{thin}
\fmfleft{l1}
\fmfright{r1,r2}
\fmf{#2}{l1,i1}
\fmfcyclen{#1,right=0.5}{i}{3}
\fmf{#3}{i2,r1}
\fmf{#4}{i3,r2}
\end{fmfgraph}
}

\newcommand{\thvlooplabel}[8]{%
\begin{fmfgraph*}(3,2)
\fmfpen{thin}
\fmfleft{l1}
\fmfright{r1,r2}
\fmf{#2,label=$#6$,l.side=left}{l1,i1}
\fmfcyclen{#1,right=0.5,label=$#5$}{i}{3}
\fmf{#3,label=$#7$,l.side=left}{i2,r1}
\fmf{#4,label=$#8$,l.side=left}{i3,r2}
\end{fmfgraph*}
}

\newcommand{\thvlooprlabel}[8]{%
\begin{fmfgraph*}(3,2)
\fmfpen{thin}
\fmfleft{l1}
\fmfright{r1,r2}
\fmf{#2,label=$#6$,l.side=left}{l1,i1}
\fmfrcyclen{#1,left=0.5,label=$#5$}{i}{3}
\fmf{#3,label=$#7$,l.side=left}{i2,r1}
\fmf{#4,label=$#8$,l.side=left}{i3,r2}
\end{fmfgraph*}
}

\newcommand{\twvloopout}[4]{%
\begin{fmfgraph}(3,2)
\fmfpen{thin}
\fmfleft{l1}
\fmfright{r1,r2}
\fmf{#2}{l1,i1}
\fmfcyclen{#1,right}{i}{2}
\fmf{#3}{i2,r1}
\fmf{#4}{i2,r2}
\end{fmfgraph}
}

\newcommand{\twvloopin}[4]{%
\begin{fmfgraph}(3,2)
\fmfpen{thin}
\fmfleft{l1}
\fmfright{r1,r2}
\fmf{#2}{l1,i1}
\fmfcyclen{#1,right,tension=0.8}{i}{2}
\fmf{#3}{i1,r1}
\fmf{#4}{i2,r2}
\end{fmfgraph}
}

\newcommand{\twvloopvout}[4]{%
\begin{fmfgraph}(3,2)
\fmfpen{thin}
\fmfleft{l1}
\fmfright{r1,r2}
\fmf{#2}{l1,i1}
\fmfcyclen{#1,right,tension=0.6}{i}{2}
\fmf{#3}{i2,i3,r1}
\fmf{#4}{i3,r2}
\end{fmfgraph}
}

\newcommand{\twvloopvin}[4]{%
\begin{fmfgraph}(3,2)
\fmfpen{thin}
\fmfleft{l1}
\fmfright{r1,r2}
\fmf{#2}{l1,i3,i1}
\fmf{#3}{r2,i3}
\fmfcyclen{#1,right,tension=0.6}{i}{2}
\fmf{#4}{i2,r1}
\end{fmfgraph}
}

\newcommand{\threeendcounterterm}[3]{%
\begin{fmfgraph}(3,2)
\fmfpen{thin}
\fmfleft{l}
\fmfright{r1,r2}
\fmfpen{thick}
\fmfv{decor.shape=cross}{i}
\fmfpen{thin}
\fmf{#1}{l,i}
\fmf{#2}{i,r2}
\fmf{#3}{i,r1}
\end{fmfgraph}
}

\newcommand{\twoendcountertermvertex}[3]{%
\begin{fmfgraph}(3,2)
\fmfpen{thin}
\fmfleft{l}
\fmfright{r2,r1}
\fmfpen{thick}
\fmfv{decor.shape=cross}{i1}
\fmfpen{thin}
\fmf{#1}{l,i1}
\fmf{#2}{i1,i2}
\fmf{#2}{i2,r1}
\fmf{#3}{i2,r2}
\end{fmfgraph}
}

\newcommand{\twoendvertexcounterterm}[3]{%
\begin{fmfgraph}(3,2)
\fmfpen{thin}
\fmfleft{l}
\fmfright{r2,r1}
\fmfpen{thick}
\fmfv{decor.shape=cross}{i2}
\fmfpen{thin}
\fmf{#1}{l,i1}
\fmf{#1}{i1,i2}
\fmf{#2}{i2,r1}
\fmf{#3}{i1,r2}
\end{fmfgraph}
}

\newcommand{\onvloop}[4]{%
\begin{fmfgraph}(3,2)
\fmfpen{thin}
\fmfleft{l1}
\fmfright{r1,r2}
\fmf{#2}{l1,i1}
\fmfcyclen{#1}{i}{1}
\fmf{#3}{i1,r1}
\fmf{#4}{i1,r2}
\end{fmfgraph}
}

\newcommand{\onvloopsplitout}[4]{%
\begin{fmfgraph}(3,2)
\fmfpen{thin}
\fmfleft{l1}
\fmfright{r1,r2}
\fmf{#2}{l1,i1}
\fmfcyclen{#1,tension=0.8}{i}{1}
\fmf{#3}{i1,i2,r1}
\fmf{#4}{i2,r2}
\end{fmfgraph}
}

\newcommand{\onvloopsplitin}[4]{%
\begin{fmfgraph}(3,2)
\fmfpen{thin}
\fmfleft{l1}
\fmfright{r1,r2}
\fmf{#2}{l1,i2,i1}
\fmfcyclen{#1,tension=0.6}{i}{1}
\fmf{#3}{i1,r1}
\fmf{#4}{i2,r2}
\end{fmfgraph}
}

\begin{document}
\begin{fmffile}{zbosondecaymffile}
%%%%%%%%%%%%%%%%%%%%%%%%%%%%%%%%%%%%%%%%%%%
% More feynmf stuff

\setlength{\unitlength}{1.8cm}

% define Z and W boson lines so that they are distinguishable in code

% define zboson option
\fmfcmd{ %
   style_def zboson expr p =
     cdraw (zigzag p);
   enddef;}

%define wboson option
\fmfcmd{ %
   style_def wboson expr p =
     cdraw (zigzag p);
   enddef;}

% define photon with arrow option
\fmfcmd{ %
   style_def photon_arrw expr p =
     cdraw (wiggly p);
     cfill (arrow p);
   enddef;}

% define z boson with arrow option
\fmfcmd{ %
   style_def zboson_arrw expr p =
     cdraw (zigzag p);
     cfill (arrow p);
   enddef;}

% define w boson with arrow option
\fmfcmd{ %
   style_def wboson_arrw expr p =
     cdraw (zigzag p);
     cfill (arrow p);
   enddef;}

% End more feynmf stuff
%%%%%%%%%%%%%%%%%%%%%%%%%%%%%%%%%%%%%%%%%%

\section{Introduction}

Nieves and Pal~\cite{NiP2005} have considered the decay of a Z~boson
to a photon and a standard four-dimensional graviton. In the ADD
scenario~\cite{ArDD1998}\footnote{The idea of large extra dimensions
  was also considered \cite{An1990,AnB1994,AnBQ1994} prior to the work of Arkani-Hamed, Dimopoulos and Dvali.}, the
graviton may be viewed from a four-dimensional perspective as gaining
a ``tower'' of massive Kaluza-Klein (KK)
excitations~\cite{GiRW1998,HaLZ1998} (this tower takes a relatively simple
form if we assume that the extra dimensions are toroidally
compactified with the copies of $S^1$ having a common radius $R/2\pi$,
with $R$ the common circumference). Although the decays of a Z~boson
involving
real production of a KK graviton excitation with a photon will be
suppressed by a gravitational coupling, the existence of a ``tower''
of particles to which the Z~boson can decay may counteract this
suppression. Because the KK excitations only couple with gravitational
strength to Standard Model particles, they will almost certainly pass
through a detector (their detection is a next-to-leading-order process in the
gravitational coupling). The ADD scenario therefore predicts that we
should see an
increase in the decay width
of the Z~boson to a photon and missing energy relative to the Standard
Model prediction~\cite{MaO1978,GaGR1979,BaRS1981,JaKW1998}.

In this paper, we calculate, to leading order, each of the decay widths of the Z~boson to a photon and a KK
graviton excitation. There are two relevant towers of KK excitations:
a spin-2 tower and a spin-0 tower. We combine the calculated widths to obtain an overall
decay width for the decay of a Z~boson to a photon and some KK graviton
excitation. This calculated width will allow the determination of
bounds on the size of large extra dimensions
when combined with experimental data on Z~decay~\cite{AlSfuture}. (The
current upper limit on the branching ratio for $Z\to \gamma + X$, with
$X$ some beyond-Standard Model invisible particle or particles, is
$O(10^{-6})$~\cite{Acetal1997}. With a ``Giga-Z'' collider setup,
this could potentially be reduced to around $O(10^{-9})$.) The amplitudes calculated are also relevant to processes in
other extra dimensional models where the Standard Model fields are
confined to a 4-brane (e.g.~the Randall-Sundrum~1 (RS1) model \cite{RaS1999}).

Current experimental limits on the size of ADD extra dimensions from
processes other than Z~boson decay come from inverse
square law experiments~\cite{Hoetal2004,Ad2002}, from consideration of
the channel $e^+ + e^- \to \gamma + {\rm graviton}$ at the LEP
experiments~\cite{lep2004}, and from consideration of the channel $p +
\bar{p} \to {\rm jet} + {\rm graviton}$ at the Tevatron
experiments~\cite{Abetal2006,Abetal2003}. In the near future, the most
likely improvement in these experimental bounds should come with the
publication of results from the D\O\ experiment using Run~II of the
Tevatron, and combination of those results with the already published
CDF Run~II bounds~\cite{Abetal2006}. Further into the future, we can
expect investigation of large extra dimensions at the Large Hadron Collider.

For the decays considered in this paper, the leading order process is a one-loop
process, because the tree-level vertex for Z~boson decay into a photon
and a graviton is absent. This means that there is a higher
order prefactor in coupling constants of the decay width when this
decay mode is compared to other, tree-level, graviton production modes
considered previously~\cite{GiRW1998,HaLZ1998,MiPP1998}. However,
there is a
large amount of experimental data for Z~boson decay, which makes
reasonable a
comparison of bounds set by this novel production process with bounds set by
other processes.

The tree-level vertex is absent because it is
derived by considering a perturbative expansion of the metric about
flat space, so that the Z-$\gamma$-graviton term comes from a
perturbation of the Z-$\gamma$ Lagrangian term, which is zero. (In
essence, each vertex involving a KK graviton excitation is derived by
``hanging'' a graviton off a propagator or an existing
vertex~\cite{NiP1998,Ve1975}.) Because we are working with bare
parameters, we are in effect working with a basis for the electroweak
sector in which there is Z-$\gamma$ mixing at one-loop level, but this
will only enter the Lagrangian via counterterm corrections. It is
clear that the amplitude calculated will not contain terms associated
with Z-$\gamma$-graviton tree-level mixing, since we could calculate
the amplitude using renormalized parameters instead of counterterms,
whereby we should have explicitly no Z-$\gamma$-graviton mixing.

At leading order, the decay is either into a
spin-2 KK excitation or into a spin-0 KK excitation (the spin-1 KK
excitation does not couple directly to matter~\cite{HaLZ1998}). The
spin-2 case is almost identical to the case of Z~boson decay to photon
plus graviton
without extra dimensions, which has been considered by Nieves and
Pal~\cite{NiP2005}. We repeat in this paper some of the detail, for
sake of completeness. The methodology of the spin-0 case is strongly
motivated by that of the spin-2 case.

In many tree-level decays
involving particles with small masses, the contribution to the decay
width of channels involving a spin-0 KK graviton
can be ignored, as the vertices involving the spin-0 KK graviton
contain the masses of the other particles (and in some cases also
contain momentum terms
that are zero on-shell). However, the
possibility of massive particles in the loop means that there is a
non-negligible contribution to the amplitude in the one-loop
calculation from the spin-0 decay channel.

For the decays into a spin-2 KK graviton and for the
decays into a spin-0 KK graviton, three types of
diagram must be considered. The first two types of diagram are those
with a fermion in the loop and those with a W~boson in the loop. (In
principle, the second type of diagram also includes diagrams with
Goldstone bosons and Fadeev-Popov ghosts in the loop, but we adopt the
unitary gauge throughout.) A
third type of diagram is required if we are to work using bare
parameters, namely the diagrams containing a counterterm. It will turn
out that the counterterm diagrams evaluate to zero, which is why we
work with bare parameters (and not renormalized parameters). The result that these diagrams evaluate to zero
supports the conclusion in~\cite{NiP2005} that such diagrams need not
be considered in the near-flat Standard Model scenario, and supports
what appears to be a ``miraculous cancellation'' of divergent terms in
the spin-0 KK graviton calculation in this paper.

We also make use of the argument of~\cite{NiP2005} that, in the spin-2
KK graviton case, conservation
of the electromagnetic current and of the energy-momentum tensor
implies a particular general form taken by the amplitude, which
simplifies the calculation. We derive an analogous argument for the
spin-0 KK graviton case.

It is possible to estimate the form that the decay width will take
prior to a full calculation. For the decay of a Z~boson into a photon
and a graviton in $3+1$-dimensional space, Nieves and
Pal~\cite{NiP2005} give the estimate
\begin{equation}
\Gamma \sim \alpha^2 G M_Z^3
\end{equation}
by
dimensional analysis. They derive this estimate by noting that the graviton coupling introduces
into the amplitude a factor of $\kappa = \sqrt{8\pi G}$, where $G$ is
Newton's constant in four dimensions; they also note that both the
Z~coupling and the photon coupling introduce into the amplitude a
factor of $\sqrt{\alpha}$, where $\alpha$ is the fine-structure
constant. (The estimate is then obtained by noting that the Z~mass,
$M_Z$, is the only dimensionful parameter remaining in the
calculation.)

For the calculation in the ADD scenario, the factor $\alpha^2 G$
remains in the width (and $G$ is still Newton's constant in four
dimensions). However, we sum over the Kaluza-Klein excitations of the
graviton by using an integral approximation over a ``density''
$\rho$~\cite{HaLZ1998}, and this density contains a factor $R^n$,
where $R/2\pi$ is the radius of the extra dimensions and $n$ is the
number of extra dimensions. The Z~mass remains the only other
dimensionful parameter (since we have summed over all KK graviton
masses). We therefore obtain the estimate
\begin{equation}
\Gamma \sim \alpha^2 G R^n M_Z^{3+n} \, .
\end{equation}

The detailed phenomenology of our more precisely calculated results is
covered in reference~\cite{AlSfuture}.

The model of toroidal compactification of the extra dimensions with a
common compactification radius $R/2\pi$ is something of a toy
model, not least because it does not suggest a mechanism for confining
the Standard Model fields to a 4-brane, as is required for the ADD
scenario. However, the model still deserves phenomenological
investigation, firstly because the calculations involved may be useful
for understanding models where the topology of the compactified
dimensions is more complicated, and secondly because the toy model
still gives some indication of the likely consistency of the ADD
scenario with experimental observations.

This paper is organised as follows. In
Sections~\ref{sec:fermiondiagrams},~\ref{sec:wdiagrams}
and~\ref{sec:countertermdiagrams}, we state the Feynman rules required 
and give the diagrams corresponding to a fermion loop, a W boson loop
and a counterterm, respectively. In Section~\ref{sec:generalform}
we present arguments giving the general form of the amplitude in each
of the spin-2 KK graviton production and the spin-0 KK graviton
production cases. Consideration of the spin-2 case, identically to that derived by
Nieves and Pal~\cite{NiP2005}, shows that the amplitude can be
expressed in terms of one CP-even and two CP-odd
coefficients. Consideration of the
spin-0 case shows that the amplitude can be expressed in terms of one
CP-even and one CP-odd coefficient. In
Section~\ref{sec:transversality}, we demonstrate that the sets of
diagrams previously presented satisfy the relevant electromagnetic and
gravitational transversality conditions, so that it is possible to use
the general form arguments. We then proceed in
Section~\ref{sec:amplitudecalculation} to calculate the coefficients
required for an expression of the amplitudes. In
Section~\ref{sec:decaywidth}, we use the expressions for the
amplitudes to calculate decay widths for the individual KK excitation
modes, and then sum over these modes (approximating the sum by an
integral~\cite{HaLZ1998}) to obtain an overall decay width.

\subsection{General notation}

We take Greek indices to range over the four
dimensions corresponding to the Standard Model brane, and Roman indices to range over the $n$ extra (bulk) dimensions.

We work with a metric tensor linearised so that
\begin{equation}
g_{\lambda\rho} = \eta_{\lambda\rho} + 2 \kappa
(h_{\lambda\rho}+\eta_{\lambda\rho}\phi)
\end{equation}
(note that this is a hybrid of the notation of~\cite{NiP2005,Ve1975}, which
do not have the extra dimensional dilaton term, and
of~\cite{HaLZ1998}, which differs by a factor of two). With this
definition, the gravitational coupling $\kappa$ satisfies
\begin{equation}
\kappa = \sqrt{8\pi G} \, ,
\end{equation}
where $G$ is Newton's constant. The fields generated by the KK
reduction may be expanded in Fourier modes and redefined in terms of
massive fields $\htil$, $\tilde{A}_{\mu i}^{\vec{n}}$ and $\ptil$ (for
the spin-2, spin-1 and spin-0 cases respectively)~\cite{HaLZ1998}. (The spin-1 field
does not couple directly to matter and is therefore neglected from now
on as a higher order contribution.) $\vec{n}$ is a vector giving the excitation level in
each of the extra dimensions.

We note that at each mass
level (i.e.~for each distinct value of the excitation vector $\vec{n}$), there are one spin-2 KK graviton excitation $\htil$ and
$n-1$ spin-0 KK graviton excitations $\ptil$ to be
considered. The $n-1$ factor comes from noting that
the Standard Model particles couple only to the trace $\ptiln$ of the
spin-0 particles, and that one degree of freedom is lost owing to the
linear dependence of the modes $\ptil$. Equivalently, we may note that
vertices involving the spin-0 particles always contain a $\delta_{ij}$
term, and each external spin-0 particle is accompanied by an
extra-dimensional ``polarisation tensor'' $e_{ij}$, which satisfies the spin
sum identity~\cite{HaLZ1998}
\begin{equation}
\sum_{s=1}^{n(n-1)/2} e^s_{ij} e^{s*}_{i'j'} =
\frac{1}{2}P^{\vec{n}}_{ii'}P^{\vec{n}}_{jj'} +
\frac{1}{2}P^{\vec{n}}_{ij'}P^{\vec{n}}_{ji'}\, \label{eq:edspinsumids},
\end{equation}
with
\begin{equation}
P^{\vec{n}}_{ij} = \delta_{ij}-\frac{n_in_j}{\vec{n}^2}\, ,
\end{equation}
which satisfies
\begin{equation}
P^{\vec{n}}_{ij}P^{\vec{n}}_{jk} = P^{\vec{n}}_{ik}, \qquad
P^{\vec{n}}_{ii} = n-1 \, ,
\label{eq:edpolarisationids}\end{equation}
so that when we calculate the modulus-squared of the amplitude, the
terms carrying extra dimensional indices look like
\begin{equation}
\delta_{ij}\delta_{i'j'}\left(
\frac{1}{2}P^{\vec{n}}_{ii'}P^{\vec{n}}_{jj'} +
\frac{1}{2}P^{\vec{n}}_{ij'}P^{\vec{n}}_{ji'} \right) ,
\end{equation}
and, using equation~\eqref{eq:edpolarisationids}, this evaluates to $n-1$,
as expected.

We define the polarisation tensors $\polg$, $\polp$ and $\polz$ as
corresponding to
the spin-2 KK graviton excitation, the photon and the Z~boson,
respectively. The tensors satisfy
\begin{align}
\polp k_{\nu} &= 0 \,\label{eq:photonpol} , \\
\polz p_{\mu} &= 0 \, \label{eq:zpol},
\end{align}
and
\begin{equation}
\polg q_{\lambda} = 0 \, , \qquad \polg q_{\rho} = 0 \,\label{eq:gravpol} .
\end{equation}
The gravitational polarisation tensor is symmetric and traceless:
\begin{equation}
\polg = \polgr \, , \qquad \polg \eta_{\lambda\rho} =0 \, .
\end{equation}

The momenta satisfy the on-shell conditions
\begin{align}
p^2 &= M_Z^2 \,\label{eq:zmomsq} ,\\
k^2 &= 0 \,\label{eq:photonmomsq} ,\\
q^2 &= \kkm^2 \, ,
\end{align}
where $\kkm$ is the mass of the KK graviton excited to level
$\vec{n}$, given for the toroidally compactified ADD scenario by~\cite{HaLZ1998}
\begin{equation}
\kkm^2 = \frac{4\pi^2 \vec{n}^2}{R^2} \, .
\end{equation}
Four-momentum conservation ($p=k+q$) yields the on-shell identity
\begin{equation}
2k \cdot q = M_Z^2 - \kkm^2 \, .
\end{equation}

We introduce notation for the off-shell amplitudes
$F^{(h)}_{\lambda\rho\mu\nu}(q,k)$ (in the case of decay to a spin-2 KK
excitation) and $F^{(\phi)}_{\mu\nu}(q,k)$ (in the case of decay to a
spin-0 KK excitation), defined by
\begin{align}
\matel^{(h)}(q,k) &= \polgs \polps \polz F^{(h)}_{\lambda\rho\mu\nu}(q,k) \label{eq:hmatelement}\\
\intertext{and}
\matel^{(\phi )}(q,k) &= \polps \polz F^{(\phi)}_{\mu\nu}(q,k) \label{eq:pmatelement}\, ,
\end{align}
respectively.

\section{Fermion loop diagrams}\label{sec:fermiondiagrams}

%%%%%%%%%%%%%%%%%%%%%%%%%%%%%%%%%%%%%%%
% htilde fermion diagrams
\begin{figure}
%\FIGURE{
\begin{center}
\subfloat[][]{\thvlooplabel{fermion}{zboson}{dbl_wiggly}{photon}{f}{Z}{\htiln}{\gamma}\label{fig:htildefermion-a}%
}
\subfloat[][]{\thvloop{fermion}{zboson}{photon}{dbl_wiggly}\label{fig:htildefermion-b}%
}
\subfloat[][]{\twvloopout{fermion}{zboson}{dbl_wiggly}{photon}\label{fig:htildefermion-c}%
}
\subfloat[][]{\twvloopin{fermion}{zboson}{dbl_wiggly}{photon}\label{fig:htildefermion-d}%
}
\subfloat[][]{\twvloopvout{fermion}{zboson}{photon}{dbl_wiggly}\label{fig:htildefermion-e}%
}
\subfloat[][]{\twvloopvin{fermion}{zboson}{dbl_wiggly}{photon}\label{fig:htildefermion-f}%
}
\caption{1-loop diagrams for the process $Z \to \gamma +
 \htiln $ involving fermions in the loop. The spin-2 KK graviton
 excitations are represented by the double wavy lines.}
\label{fig:htildefermion}
\end{center}
%}
\end{figure}
% end htilde fermion diagrams
%%%%%%%%%%%%%%%%%%%%%%%%%%%%%%%%%%%%%%%
Figure~\ref{fig:htildefermion} contains the diagrams for the
$Z\to\gamma +\htiln$ process with a fermion in the loop, and
figure~\ref{fig:phitildefermion} contains the diagrams for the
$Z\to\gamma +\ptiln$ process with a fermion in the loop.
%%%%%%%%%%%%%%%%%%%%%%%%%%%%%%%%%%%%%%%
% phitilde fermion diagrams
\begin{figure}
%\FIGURE{
\begin{center}
\subfloat[][]{\thvlooplabel{fermion}{zboson}{double}{photon}{f}{Z}{\ptiln}{\gamma}\label{fig:phitildefermion-a}%
}
\subfloat[][]{\thvloop{fermion}{zboson}{photon}{double}\label{fig:phitildefermion-b}%
}
\subfloat[][]{\twvloopout{fermion}{zboson}{double}{photon}\label{fig:phitildefermion-c}%
}
\subfloat[][]{\twvloopin{fermion}{zboson}{double}{photon}\label{fig:phitildefermion-d}%
}
\subfloat[][]{\twvloopvout{fermion}{zboson}{photon}{double}\label{fig:phitildefermion-e}%
}
\subfloat[][]{\twvloopvin{fermion}{zboson}{double}{photon}\label{fig:phitildefermion-f}%
}
\caption{1-loop diagrams for the process $Z \to \gamma +
  \ptiln $ involving fermions in the loop. The spin-0 KK graviton
  excitations are represented by double straight lines.}
\label{fig:phitildefermion}
\end{center}
%}
\end{figure}
% end phitilde fermion diagrams
%%%%%%%%%%%%%%%%%%%%%%%%%%%%%%%%%%%%%%%
\setlength{\unitlength}{1cm}
%%%%%%%%%%%%%%%%%%%%%%%%%%%%%%%%%%%%%%
% Vertex Feynman rules for the first set of diagrams
\begin{figure}
\begin{center}
\subfloat{%
\begin{tabular}{c}
\fmfframevertex{%
\begin{fmfgraph*}(2,2)
\fmfsurround{v1,v2,v3}
\fmf{fermion}{v3,i1,v2}
\fmf{zboson}{v1,i1}
\fmfplab{Z_{\mu}}{v1}
\end{fmfgraph*}}\\
$\frac{-ig\tilde{\gamma}_{\mu}}{2 \cos \theta_W}$
%See Nieves and Pal p.7
\end{tabular}
}
\subfloat{%
\begin{tabular}{c}
\fmfframevertex{%
\begin{fmfgraph*}(2,2)
\fmfsurround{v1,v2,v3}
\fmf{fermion}{v3,i1,v2}
\fmf{photon}{v1,i1}
\fmfplab{A_{\nu}}{v1}
\end{fmfgraph*}}\\
$-ieQ_f\gamma_{\nu}$
%See Nieves and Pal p.7
\end{tabular}
}
\subfloat{%
\begin{tabular}{c}
\fmfframevertex{%
\begin{fmfgraph*}(2,2)
\fmfsurround{v1,v2,v3}
\fmf{fermion,label=$k_1$,l.side=left}{v3,i1}
\fmf{fermion,label=$k_2$,l.side=left}{i1,v2}
\fmf{dbl_wiggly}{v1,i1}
\fmfplab{\htil}{v1}
\end{fmfgraph*}}\\
%$-i \frac{\kappa}{4} ( \gamma_{\lambda}(k_{1\rho}+k_{2\rho}) +
%\gamma_{\rho}(k_{1\lambda}+k_{2\lambda}) $ \\
%$ - 2\eta_{\lambda\rho}(\slashed{k}_1+\slashed{k}_2 - 2m_f))$
$-i \kappa V_{\lambda\rho}(k_1,k_2)$
%See Han et al.
\end{tabular}
}
\subfloat{%
\begin{tabular}{c}
\fmfframevertex{%
\begin{fmfgraph*}(2,2)
\fmfleft{v4,v3}
\fmfright{v1,v2}
\fmf{fermion}{v4,i1}
\fmf{fermion}{i1,v3}
\fmf{photon}{v2,i1}
\fmf{dbl_wiggly}{v1,i1}
\fmfplab{A_{\nu}}{v2}
\fmfplab{\htil}{v1}
\end{fmfgraph*}}\\
$ieQ_f \frac{\kappa}{2} (C_{\lambda\rho\nu\sigma} -\eta_{\lambda\rho}\eta_{\nu\sigma})\gamma^{\sigma}$
%See Han et al.
\end{tabular}
}
\subfloat{%
\begin{tabular}{c}
\fmfframevertex{%
\begin{fmfgraph*}(2,2)
\fmfleft{v4,v3}
\fmfright{v1,v2}
\fmf{fermion}{v4,i1}
\fmf{fermion}{i1,v3}
\fmf{photon}{v2,i1}
\fmf{dbl_wiggly}{v1,i1}
\fmfplab{Z_{\mu}}{v2}
\fmfplab{\htil}{v1}
\end{fmfgraph*}}\\
$i\frac{g}{2 \cos \theta_W}\frac{\kappa}{2} \tilde{\gamma}^{\sigma} (C_{\lambda\rho\mu\sigma}-\eta_{\lambda\rho}\eta_{\mu\sigma})$
%See Han et al.
\end{tabular}
}
\subfloat{%
\begin{tabular}{c}
\fmfframevertex{%
\begin{fmfgraph*}(2,2)
\fmfsurround{v1,v2,v3}
\fmf{photon_arrw,label=$k_1$,l.side=left}{v3,i1}
\fmfplab{A_{\beta}}{v3}
\fmf{photon_arrw,label=$k_2$,l.side=left}{i1,v2}
\fmfplab{A_{\nu}}{v2}
\fmf{dbl_wiggly}{v1,i1}
\fmfplab{\htil}{v1}
\end{fmfgraph*}}\\
%$i\kappa [(k_1 \cdot k_2) C_{\lambda \rho \beta \nu} + D_{\lambda \rho
%    \beta \nu} (k_1,k_2) ]$
$i\kappa c_{\lambda\rho\beta\nu}(k_1,k_2)$
%See Han et al.
\end{tabular}
}
\subfloat{%
\begin{tabular}{c}
\fmfframevertex{%
\begin{fmfgraph*}(2,2)
\fmfsurround{v1,v2,v3}
\fmf{zboson_arrw,label=$k_1$,l.side=left}{v3,i1}
\fmfplab{Z_{\mu}}{v3}
\fmf{zboson_arrw,label=$k_2$,l.side=left}{i1,v2}
\fmfplab{Z_{\alpha}}{v2}
\fmf{dbl_wiggly}{v1,i1}
\fmfplab{\htil}{v1}
\end{fmfgraph*}}\\
%$i\kappa [(k_1 \cdot k_2 - m_Z^2) C_{\lambda \rho \mu \alpha} + D_{\lambda \rho
%    \mu \alpha} (k_1,k_2) ]$
$i\kappa c^Z_{\lambda \rho \mu \alpha}(k_1,k_2)$
%See Han et al.
\end{tabular}
}
\subfloat{%
\begin{tabular}{c}
\fmfframevertex{%
\begin{fmfgraph*}(2,2)
\fmfsurround{v1,v2,v3}
\fmf{fermion,label=$k_1$,l.side=left}{v3,i1}
\fmf{fermion,label=$k_2$,l.side=left}{i1,v2}
\fmf{double}{v1,i1}
\fmfplab{\ptiln}{v1}
\end{fmfgraph*}}\\
$2i\omega\kappa\left( \frac{3}{4} \slashed{k}_1 +
\frac{3}{4}\slashed{k}_2 - 2 m_f \right)$
\end{tabular}
}
\subfloat{%
\begin{tabular}{c}
\fmfframevertex{%
\begin{fmfgraph*}(2,2)
\fmfleft{v4,v3}
\fmfright{v1,v2}
\fmf{fermion}{v4,i1}
\fmf{fermion}{i1,v3}
\fmf{photon}{v2,i1}
\fmfplab{A_{\nu}}{v2}
\fmf{double}{v1,i1}
\fmfplab{\ptiln}{v1}
\end{fmfgraph*}}\\
$-3i\omega eQ_f\kappa \gamma_{\nu}$
%See Han et al.
\end{tabular}
}
\subfloat{%
\begin{tabular}{c}
\fmfframevertex{%
\begin{fmfgraph*}(2,2)
\fmfleft{v4,v3}
\fmfright{v1,v2}
\fmf{fermion}{v4,i1}
\fmf{fermion}{i1,v3}
\fmf{zboson}{v2,i1}
\fmfplab{Z_{\mu}}{v2}
\fmf{double}{v1,i1}
\fmfplab{\ptiln}{v1}
\end{fmfgraph*}}\\
$-3i\omega\frac{g}{2\cos \theta_W} \kappa \tilde{\gamma}_{\mu}$
%See Han et al.
\end{tabular}
}
\subfloat{%
\begin{tabular}{c}
\fmfframevertex{%
\begin{fmfgraph*}(2,2)
\fmfsurround{v1,v2,v3}
\fmf{photon}{v3,i1}
\fmfplab{A_{\beta}}{v3}
\fmf{photon}{i1,v2}
\fmfplab{A_{\nu}}{v2}
\fmf{double}{v1,i1}
\fmfplab{\ptiln}{v1}
\end{fmfgraph*}}\\
$0$
%See Han et al.
\end{tabular}
}
\subfloat{%
\begin{tabular}{c}
\fmfframevertex{%
\begin{fmfgraph*}(2,2)
\fmfsurround{v1,v2,v3}
\fmf{zboson}{v3,i1}
\fmfplab{Z_{\mu}}{v3}
\fmf{zboson}{i1,v2}
\fmfplab{Z_{\alpha}}{v2}
\fmf{double}{v1,i1}
\fmfplab{\ptiln}{v1}
\end{fmfgraph*}}\\
$2i\omega\kappa\eta_{\mu\alpha}M_Z^2$
%See Han et al.
\end{tabular}
}
\caption{Feynman rules for vertices in fermion loop diagrams. The
  extra-dimensional indices are omitted from the vertices involving
  $\ptil$ (each vertex gains a coefficient of $\delta_{ij}$). Symbols
  and tensors are defined in the main text. Arrows on bosonic lines
  indicate directions of flow of momenta.}
\label{fig:fermionfeynman}
\end{center}
\end{figure}
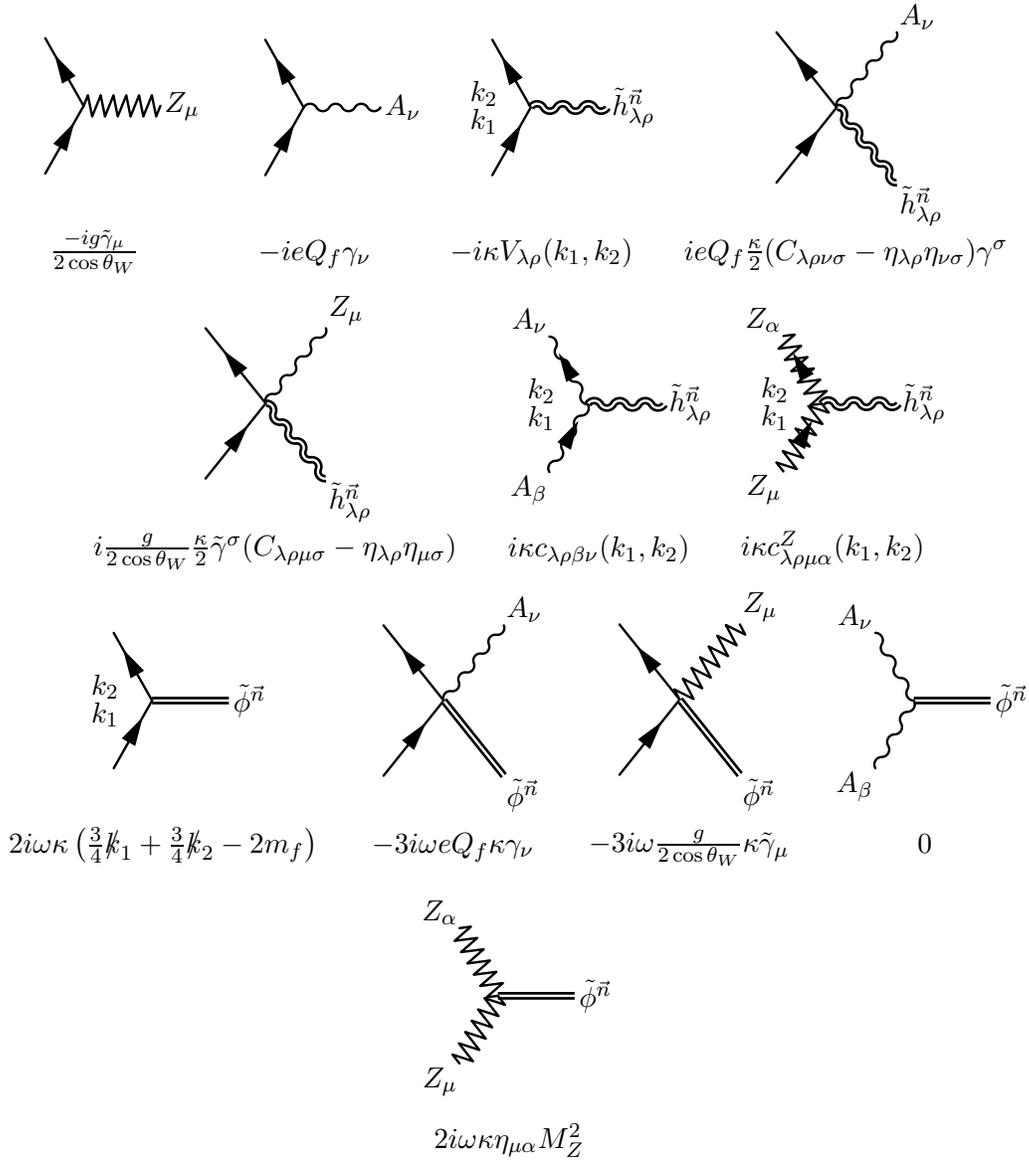
% end vertex Feynman rules for the first set of diagrams
%%%%%%%%%%%%%%%%%%%%%%%%%%%%%%%%%%%%
\setlength{\unitlength}{1.8cm}
The Feynman rules for the vertices occurring in the diagrams are given
for reference in figure~\ref{fig:fermionfeynman}. Their derivation is
given in~\cite{GiRW1998,HaLZ1998}. 
Some of the tensor contributions to the vertex factors are abbreviated
for legibility; the abbreviations used are
\begin{align}
C_{\lambda\rho\nu\sigma} &\equiv 
\eta_{\lambda\nu}\eta_{\rho\sigma} +
\eta_{\lambda\sigma}\eta_{\rho\nu} -
\eta_{\lambda\rho}\eta_{\nu\sigma} \, ,\label{eq:bigcdefine}\\
c_{\lambda\rho\beta\nu}(k_1,k_2) &\equiv  (k_1 \cdot
k_2)C_{\lambda\rho\beta\nu} + \eta_{\lambda\rho}k_{1\nu}k_{2\beta}
-\nonumber\\
&\phantom{\equiv} - 
\left[\eta_{\lambda\nu}k_{1\rho}k_{2\beta}+\eta_{\lambda\beta}k_{1\nu}k_{2\rho}-\eta_{\beta\nu}k_{1\lambda}k_{2\rho}+(\lambda
  \leftrightarrow \rho)\right] \, ,\label{eq:smallcdefine}\\
c^Z_{\lambda\rho\beta\nu}(k_1,k_2) &\equiv 
c_{\lambda\rho\beta\nu}-M_Z^2C_{\lambda\rho\beta\nu} \, ,\label{eq:smallczdefine}\\
V_{\lambda\rho}(k_1,k_2) &\equiv 
\frac{1}{4}\left[\gamma_{\lambda}(k_1+k_2)_{\rho} +
  \gamma_{\rho}(k_1+k_2)_{\lambda} -
  2\eta_{\lambda\rho}(\slashed{k}_1+\slashed{k}_2-2m_f) \right] .
\end{align}
We use $Q_f$ to denote the fermion
charge, and have written
\begin{equation}
\tilde{\gamma}_{\mu} \equiv \gamma_{\mu}(X_f + Y_f \gamma_5) \, ;
\end{equation}
$X_f$ and $Y_f$ are scalars.
We have also defined (as in~\cite{HaLZ1998})
\begin{equation}
\omega \equiv \sqrt{\frac{2}{3(n+2)}}\, .
\end{equation}

Writing $F_{\lambda\rho\mu\nu}^{(hf)}$ and
$F_{\mu\nu}^{(\phi f)}$ to denote the contributions from
the fermion loop diagrams to the off-shell amplitudes
$F_{\lambda\rho\mu\nu}^{(h)}$ and $F_{\mu\nu}^{(\phi )}$,
respectively, and absorbing constants by defining
\begin{align}
F_{\lambda\rho\mu\nu}^{(hf)} & \equiv \frac{\kappa e Q_f g}{2 \cos
  \theta_W}\allconthf \\
\intertext{and}
F_{\mu\nu}^{(\phi f)} & \equiv \frac{\kappa e Q_f g}{2 \cos
  \theta_W}\allcontpf \, ,
\end{align}
we may write the contributions to the amplitudes from individual
diagrams as
\begin{align}
\diagconthf{a} &= -\linttr S(l-k) \gamma_{\nu} S(l)
V_{\lambda\rho}(l,l+q) S(l+q) \right] \label{eq:diagconthfa},\\
\diagconthf{b} &= -\linttr S(l-q) V_{\lambda\rho}(l,l-q) S(l)
  \gamma_{\nu} S(l+k) \right] \label{eq:diagconthfb},\\
\diagconthf{c} &= -i a_{\lambda\rho\nu\sigma}
\Pi_{\mu}^{\phantom{\mu}\sigma}(p) \label{eq:diagconthfc}\, ,\\
\diagconthf{d} &= -i a_{\lambda\rho\mu\sigma}\Pi^{\sigma}_{\phantom{\sigma}\nu}(k) \label{eq:diagconthfd}\, ,\\
\diagconthf{e} &=
D^{\alpha\beta}(p)c_{\lambda\rho\beta\nu}(p,k)\Pi_{\mu\alpha}(p) \label{eq:diagconthfe}\,
,\\
\diagconthf{f} &=
D_Z^{\alpha\beta}(k)c^Z_{\lambda\rho\mu\alpha}(p,k)\Pi_{\beta\nu}(k)
\label{eq:diagconthff}\, ,\\
\intertext{and}
\diagcontpf{a} &= 2\omega \linttr S(l-k)\gamma_{\nu}S(l)\left(
\frac{3}{2}\slashed{l} + \frac{3}{4}\slashed{q} - 2m_f \right)S(l+q)
\right]  \label{eq:diagcontpfa},\\
\diagcontpf{b} &= 2\omega \linttr S(l-q) \left( \frac{3}{2}\slashed{l}
- \frac{3}{4}\slashed{q} - 2m_f \right) S(l) \gamma_{\nu}S(l+k)
\right]  \label{eq:diagcontpfb},\\
\diagcontpf{c} &= -3i\omega\Pi_{\mu\nu}(p) \label{eq:diagcontpfc}\, ,\\
\diagcontpf{d} &= -3i\omega\Pi_{\mu\nu}(k) \label{eq:diagcontpfd}\, ,\\
\diagcontpf{e} &= 0 \label{eq:diagcontpfe}\, ,\\
\diagcontpf{f} &= 2\omega\eta_{\mu\alpha}M_Z^2
D_Z^{\alpha\beta}(k)\Pi_{\beta\nu}(k) \label{eq:diagcontpff}\, ,\\
\intertext{where we have written}
a_{\lambda\rho\mu\nu} &\equiv \eta_{\lambda\rho}\eta_{\mu\nu} -
\frac{1}{2}\eta_{\lambda\mu}\eta_{\rho\nu} -
\frac{1}{2}\eta_{\lambda\nu}\eta_{\rho\mu}\\
\intertext{and}
\Pi_{\mu\nu}(k) &\equiv \linttr S(l) \gamma_{\nu} S(l+k) \right] \label{eq:bigpiintegraldefine}
  ,\\
\intertext{and have defined the propagators}
S(l) &\equiv \frac{i(\slashed{l}+m_f)}{l^2-m_f^2} \, ,\\
D^{\alpha\beta}(p) &\equiv \frac{-i\eta^{\alpha\beta}}{p^2} \, ,\\
D_Z^{\alpha\beta}(k) &\equiv
\frac{i}{k^2-M_Z^2}\left(-\eta^{\alpha\beta}+\frac{k^{\alpha}k^{\beta}}{M_Z^2}\right) \label{eq:zpropagatordefine}
\end{align}
for the fermion, photon and Z~boson respectively (note that these are
different from the definitions used in~\cite{NiP2005}). This amounts
to a choice of unitary gauge for the electroweak sector. (We shall
later need the propagator for the W~boson, which we take to be that
for the Z~boson with $M_Z \to M_W$.)

\section{W loop diagrams}\label{sec:wdiagrams}

\setlength{\unitlength}{1.45cm}
%%%%%%%%%%%%%%%%%%%%%%%%%%%%%%%%%%%%%%%%%%%%%%%%%%
% htilde W loop diagrams
\begin{figure}[!hbp]
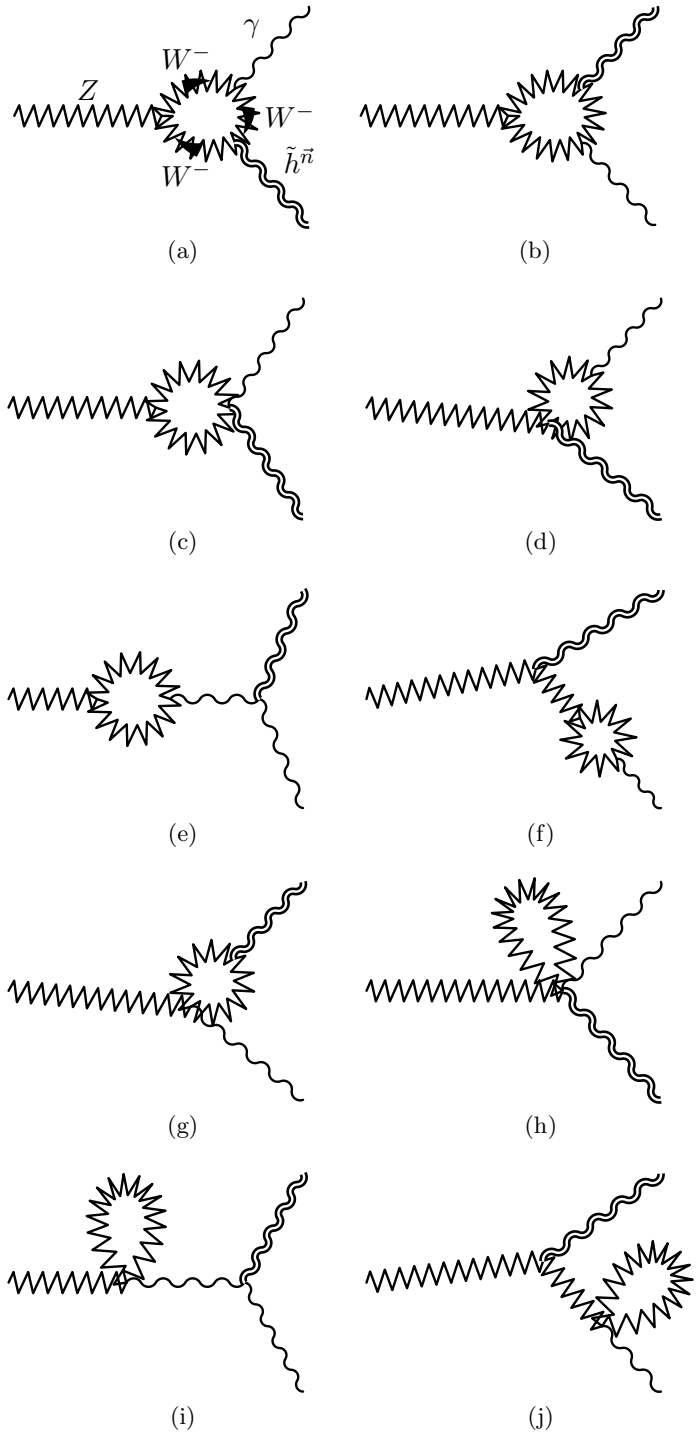

%\FIGURE{
\begin{center}
\subfloat[][]{\thvlooprlabel{wboson_arrw}{zboson}{dbl_wiggly}{photon}{W^-}{Z}{\htiln}{\gamma}}
\subfloat[][]{\thvloop{wboson}{zboson}{photon}{dbl_wiggly}}\\
\subfloat[][]{\twvloopout{wboson}{zboson}{dbl_wiggly}{photon}}
\subfloat[][]{\twvloopin{wboson}{zboson}{dbl_wiggly}{photon}}\\
\subfloat[][]{\twvloopvout{wboson}{zboson}{photon}{dbl_wiggly}}
\subfloat[][]{\twvloopvin{wboson}{zboson}{dbl_wiggly}{photon}}\\
\subfloat[][]{\twvloopin{wboson}{zboson}{photon}{dbl_wiggly}}
\subfloat[][]{\onvloop{wboson}{zboson}{dbl_wiggly}{photon}}\\
\subfloat[][]{\onvloopsplitout{wboson}{zboson}{photon}{dbl_wiggly}}
\subfloat[][]{\onvloopsplitin{wboson}{zboson}{photon}{dbl_wiggly}}
\caption{1-loop diagrams for the process $Z \to \gamma +
  \htiln$ involving W bosons in the loop. We note that once one has
  defined a convention for charge and momentum flow in the loop,
  diagrams (a) and (b) must be treated as distinct.}
\label{fig:htildewboson}
\end{center}
%}
\end{figure}
% end htilde W loop diagrams
%%%%%%%%%%%%%%%%%%%%%%%%%%%%%%%%%%%%%%%%%%%%%%%%%%
Figure~\ref{fig:htildewboson} contains the diagrams for the
$Z\to\gamma +\htiln$ process with a W~boson in the loop, and
figure~\ref{fig:phitildewboson} contains the diagrams for the
$Z\to\gamma +\ptiln$ process with a W~boson in the loop.%
\setlength{\unitlength}{1.55cm}
%%%%%%%%%%%%%%%%%%%%%%%%%%%%%%%%%%%%%%%%%%%%%%%%%%
% phitilde W loop diagrams
\begin{figure}[!hbp]
\begin{center}
\subfloat[][]{\thvlooprlabel{wboson_arrw}{zboson}{double}{photon}{W^-}{Z}{\ptiln}{\gamma}}
\subfloat[][]{\thvloop{wboson}{zboson}{photon}{double}}\\
\subfloat[][]{\twvloopout{wboson}{zboson}{double}{photon}}
\subfloat[][]{\twvloopin{wboson}{zboson}{double}{photon}}\\
\subfloat[][]{\twvloopvout{wboson}{zboson}{photon}{double}}
\subfloat[][]{\twvloopvin{wboson}{zboson}{double}{photon}}\\
\subfloat[][]{\twvloopin{wboson}{zboson}{photon}{double}}
\subfloat[][]{\onvloop{wboson}{zboson}{double}{photon}}\\
\subfloat[][]{\onvloopsplitout{wboson}{zboson}{photon}{double}}
\subfloat[][]{\onvloopsplitin{wboson}{zboson}{photon}{double}}
\caption{1-loop diagrams for the process $Z \to \gamma +
  \ptiln$ involving W bosons in the loop.}
\label{fig:phitildewboson}
\end{center}
\end{figure}%
% end phitilde W loop diagrams
%%%%%%%%%%%%%%%%%%%%%%%%%%%%%%%%%%%%%%%%%%%%%%%%%%
\setlength{\unitlength}{1cm}%
%%%%%%%%%%%%%%%%%%%%%%%%%%%%%%%%%%%%%%
% Extra vertex Feynman rules for the W boson loop diagrams
\begin{figure}
%\FIGURE{
\begin{center}
\subfloat{%
\begin{tabular}{c}
\fmfframevertex{%
\begin{fmfgraph*}(2,2)
\fmfsurround{v1,v2,v3}
\fmf{wboson_arrw,label=$k_1$,l.side=left}{v3,i1}
\fmfplab{W^-_{\alpha}}{v3}
\fmf{wboson_arrw,label=$k_2$,l.side=right}{v2,i1}
\fmfplab{W^+_{\beta}}{v2}
\fmf{zboson_arrw,label=$k_3$,l.side=right}{v1,i1}
\fmfplab{Z_{\lambda}}{v1}
\end{fmfgraph*}}\\
$- i g \cos \theta_W N_{\alpha\beta\lambda}(k_1,k_2,k_3)$
%See Nieves and Pal p.15
\end{tabular}
}
\subfloat{%
\begin{tabular}{c}
\fmfframevertex{%
\begin{fmfgraph*}(2,2)
\fmfsurround{v1,v2,v3}
\fmf{wboson_arrw,label=$k_1$,l.side=left}{v3,i1}
\fmfplab{W^-_{\alpha}}{v3}
\fmf{wboson_arrw,label=$k_2$,l.side=right}{v2,i1}
\fmfplab{W^+_{\beta}}{v2}
\fmf{photon_arrw,label=$k_3$,l.side=right}{v1,i1}
\fmfplab{A_{\lambda}}{v1}
\end{fmfgraph*}}\\
$- i e N_{\alpha\beta\lambda}(k_1,k_2,k_3)$
%See Nieves and Pal p.15
\end{tabular}
}
\subfloat{%
\begin{tabular}{c}
\fmfframevertex{%
\begin{fmfgraph*}(2,2)
\fmfsurround{v1,v2,v3}
\fmf{wboson_arrw,label=$k_1$,l.side=left}{v3,i1}
\fmfplab{W^-_{\alpha}}{v3}
\fmf{wboson_arrw,label=$k_2$,l.side=right}{v2,i1}
\fmfplab{W^+_{\beta}}{v2}
\fmf{dbl_wiggly}{v1,i1}
\fmfplab{\htil}{v1}
\end{fmfgraph*}}\\
%$-i \frac{\kappa}{4} ( \gamma_{\lambda}(k_{1\rho}+k_{2\rho}) +
%\gamma_{\rho}(k_{1\lambda}+k_{2\lambda}) $ \\
%$ - 2\eta_{\lambda\rho}(\slashed{k}_1+\slashed{k}_2 - 2m_f))$
$i \kappa c^W_{\lambda\rho\alpha\beta}(k_1,-k_2)$
%See Han et al.
\end{tabular}
}
\subfloat{%
\begin{tabular}{c}
\fmfframevertex{%
\begin{fmfgraph*}(2,2)
\fmfleft{v4,v3}
\fmfright{v1,v2}
\fmf{wboson_arrw,label=$k_1$,l.side=left}{v4,i1}
\fmfplab{W^-_{\alpha}}{v4}
\fmf{wboson_arrw,label=$k_2$,l.side=right}{v3,i1}
\fmfplab{W^+_{\beta}}{v3}
\fmf{photon_arrw,label=$k_3$,l.side=right}{i1,v2}
\fmfplab{A_{\nu}}{v2}
\fmf{dbl_wiggly}{v1,i1}
\fmfplab{\htil}{v1}
\end{fmfgraph*}}\\
$-ie\kappa d_{\lambda\rho\alpha\beta\nu}(k_1,k_2,-k_3)$
%See Han et al.
\end{tabular}
}
\subfloat{%
\begin{tabular}{c}
\fmfframevertex{%
\begin{fmfgraph*}(2,2)
\fmfleft{v4,v3}
\fmfright{v1,v2}
\fmf{wboson_arrw,label=$k_1$,l.side=left}{v4,i1}
\fmfplab{W^-_{\alpha}}{v4}
\fmf{wboson_arrw,label=$k_2$,l.side=right}{v3,i1}
\fmfplab{W^+_{\beta}}{v3}
\fmf{zboson_arrw,label=$k_3$,l.side=left}{v2,i1}
\fmfplab{Z_{\mu}}{v2}
\fmf{dbl_wiggly}{v1,i1}
\fmfplab{\htil}{v1}
\end{fmfgraph*}}\\
$-ig \cos \theta_W \kappa d_{\lambda\rho\alpha\beta\mu}(k_1,k_2,k_3)$
%See Han et al.
\end{tabular}
}
\subfloat{%
\begin{tabular}{c}
\fmfframevertex{%
\begin{fmfgraph*}(2,2)
\fmfleft{v4,v3}
\fmfright{v1,v2}
\fmf{wboson}{v4,i1}
\fmfplab{W^-_{\alpha}}{v4}
\fmf{wboson}{v3,i1}
\fmfplab{W^+_{\beta}}{v3}
\fmf{zboson}{v2,i1}
\fmfplab{Z_{\mu}}{v2}
\fmf{photon}{v1,i1}
\fmfplab{A_{\nu}}{v1}
\end{fmfgraph*}}\\
$-ieg \cos \theta_W R_{\alpha\beta\mu\nu}$
%See Nieves and Pal
\end{tabular}
}
\subfloat{%
\begin{tabular}{c}
\fmfframevertex{%
\begin{fmfgraph*}(2,2)
\fmfsurround{v1,v2,v3,v4,v5}
\fmf{photon}{v1,i1}
\fmfplab{A_{\nu}}{v1}
\fmf{wboson}{v2,i1}
\fmfplab{W^+_{\beta}}{v2}
\fmf{wboson}{v3,i1}
\fmfplab{W^-_{\alpha}}{v3}
\fmf{dbl_wiggly}{v4,i1}
\fmfplab{\htil}{v4}
\fmf{zboson}{v5,i1}
\fmfplab{Z_{\mu}}{v5}
\end{fmfgraph*}}\\
$-ieg \cos \theta_W \kappa R_{\lambda\rho\alpha\beta\mu\nu} $
%See Han et al.
\end{tabular}
}
\subfloat{%
\begin{tabular}{c}
\fmfframevertex{%
\begin{fmfgraph*}(2,2)
\fmfsurround{v1,v2,v3}
\fmf{wboson_arrw}{v3,i1}
\fmfplab{W^-_{\alpha}}{v3}
\fmf{wboson_arrw}{v2,i1}
\fmfplab{W^+_{\beta}}{v2}
\fmf{double}{v1,i1}
\fmfplab{\ptiln}{v1}
\end{fmfgraph*}}\\
$2i\omega\kappa\eta_{\alpha\beta}M_W^2$
%See Han et al.
\end{tabular}
}
\subfloat{%
\begin{tabular}{c}
\fmfframevertex{%
\begin{fmfgraph*}(2,2)
\fmfleft{v4,v3}
\fmfright{v1,v2}
\fmf{wboson_arrw}{v4,i1}
\fmfplab{W^-_{\alpha}}{v4}
\fmf{wboson_arrw}{v3,i1}
\fmfplab{W^+_{\beta}}{v3}
\fmf{photon_arrw}{i1,v2}
\fmfplab{A_{\nu}}{v2}
\fmf{double}{v1,i1}
\fmfplab{\ptiln}{v1}
\end{fmfgraph*}}\\
\rule{2cm}{0pt}$0$\rule{2cm}{0pt}
%See Han et al.
\end{tabular}
}
\subfloat{%
\begin{tabular}{c}
\fmfframevertex{%
\begin{fmfgraph*}(2,2)
\fmfleft{v4,v3}
\fmfright{v1,v2}
\fmf{wboson_arrw}{v4,i1}
\fmfplab{W^-_{\alpha}}{v4}
\fmf{wboson_arrw}{v3,i1}
\fmfplab{W^+_{\beta}}{v3}
\fmf{zboson_arrw}{v2,i1}
\fmfplab{Z_{\mu}}{v2}
\fmf{double}{v1,i1}
\fmfplab{\ptiln}{v1}
\end{fmfgraph*}}\\
\rule{2cm}{0pt}$0$\rule{2cm}{0pt}
%See Han et al.
\end{tabular}
}
\subfloat{%
\begin{tabular}{c}
\fmfframevertex{%
\begin{fmfgraph*}(2,2)
\fmfsurround{v1,v2,v3,v4,v5}
\fmf{photon}{v1,i1}
\fmfplab{A_{\nu}}{v1}
\fmf{wboson}{v2,i1}
\fmfplab{W^+_{\beta}}{v2}
\fmf{wboson}{v3,i1}
\fmfplab{W^-_{\alpha}}{v3}
\fmf{double}{v4,i1}
\fmfplab{\ptiln}{v4}
\fmf{zboson}{v5,i1}
\fmfplab{Z_{\mu}}{v5}
\end{fmfgraph*}}\\
$0$
%See Han et al.
\end{tabular}
}

\caption{Extra Feynman rules for vertices in diagrams with W boson
  loops. Additional vertex functions are defined in the main text.}
\label{fig:wfeynman}
\end{center}
%}
\end{figure}
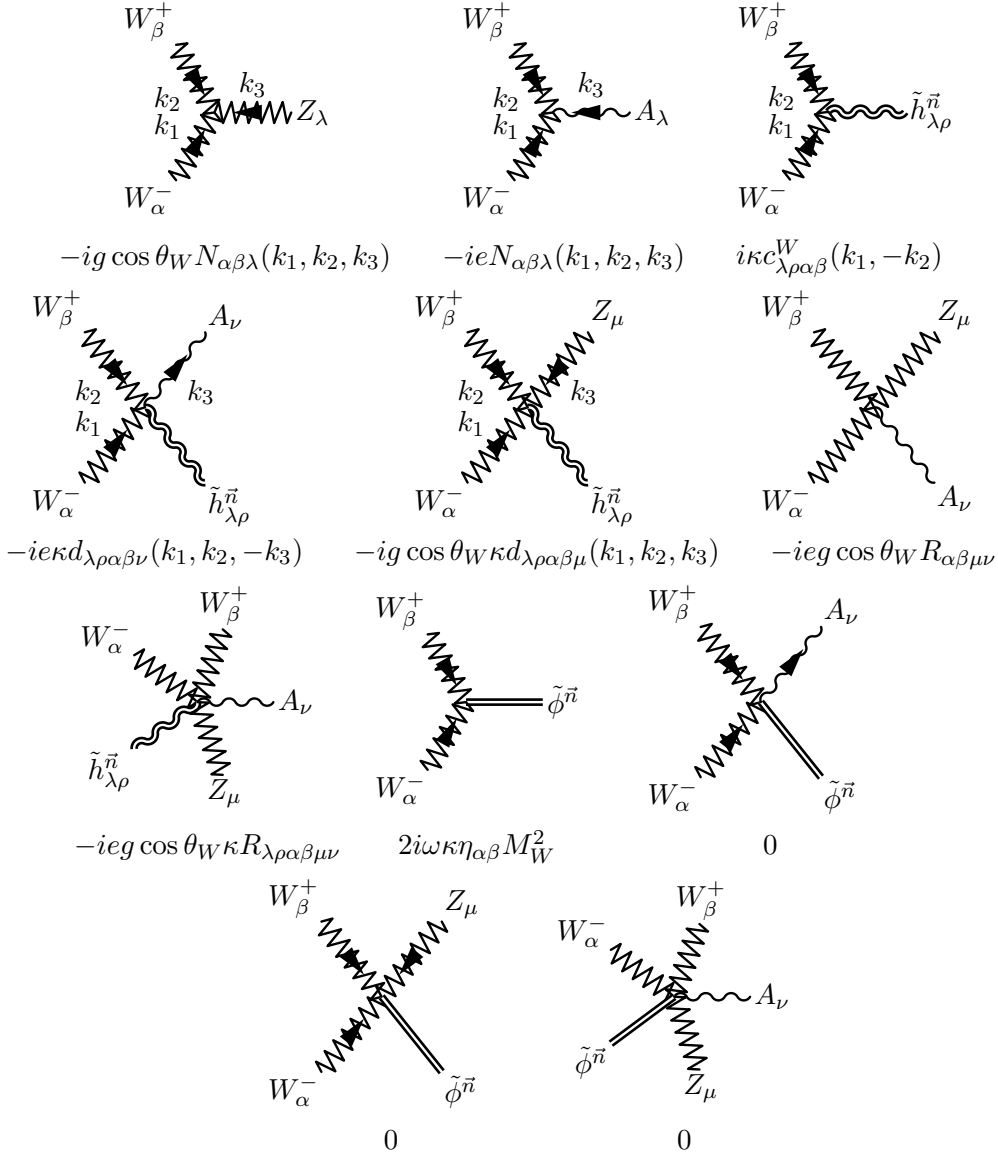%
% end extra vertex Feynman rules for the W boson loop diagrams
%%%%%%%%%%%%%%%%%%%%%%%%%%%%%%%%%%%%%%%%%%%
\setlength{\unitlength}{1.8cm}%
The Feynman rules for the vertices occurring in these diagrams additional to
the fermionic loop case are given
for reference in figure~\ref{fig:wfeynman}. Their derivation is given
in reference~\cite{HaLZ1998}. Again, we have abbreviated for legibility some of
the tensor contributions to the vertex factors; the additional
abbreviations used are
\begin{align}
N_{\alpha\beta\gamma}(k_1,k_2,k_3) &\equiv 
\eta_{\beta\gamma}(k_3-k_2)_{\alpha} +
\eta_{\gamma\alpha}(k_1-k_3)_{\beta}+\eta_{\alpha\beta}(k_2-k_1)_{\gamma}
\, ,\\
c^W_{\lambda\rho\beta\nu}(k_1,k_2) &\equiv  c_{\lambda\rho\beta\nu} - M_W^2
C_{\lambda\rho\beta\nu} \, ,\\
d_{\lambda\rho\alpha\beta\nu}(k_1,k_2,k_3) &\equiv 
C_{\lambda\rho\alpha\beta}(k_1-k_2)_{\nu} +
C_{\lambda\rho\alpha\nu}(k_3-k_1)_{\beta} +
C_{\lambda\rho\beta\nu}(k_2-k_3)_{\alpha} + \nonumber\\
 &\phantom{\equiv}+[\eta_{\lambda\alpha}\eta_{\beta\nu}(k_2-k_3)_{\rho} +
  \eta_{\lambda\beta}\eta_{\alpha\nu}(k_3-k_1)_{\rho} + \nonumber\\
 &\phantom{\equiv+[} + \eta_{\lambda\nu}\eta_{\alpha\beta}(k_1-k_2)_{\rho} + (\lambda
  \leftrightarrow \rho)] \, ,\\
R_{\alpha\beta\mu\nu} &\equiv  2\eta_{\alpha\beta}\eta_{\mu\nu} -
\eta_{\alpha\mu}\eta_{\beta\nu} - \eta_{\alpha\nu}\eta_{\beta\mu} \,
,\\
R_{\lambda\rho\alpha\beta\mu\nu} &\equiv 
\eta_{\lambda\rho}R_{\alpha\beta\nu\mu} -
\eta_{\lambda\alpha}R_{\rho\beta\nu\mu} -
\eta_{\lambda\beta}R_{\rho\alpha\nu\mu} -
\eta_{\lambda\nu}R_{\rho\mu\alpha\beta} -
\eta_{\lambda\mu}R_{\rho\nu\alpha\beta} \, .
\end{align}
%\clearpage
Writing $F_{\lambda\rho\mu\nu}^{(hW)}$ and
$F_{\mu\nu}^{(\phi W)}$ to denote the contributions from
the W~boson loop diagrams to the off-shell amplitudes
$F_{\lambda\rho\mu\nu}^{(h)}$ and $F_{\mu\nu}^{(\phi )}$,
respectively, and this time absorbing constants by defining
\begin{align}
F_{\lambda\rho\mu\nu}^{(hW)} & \equiv \kappa e g \cos
  \theta_W \allconthw \\
\intertext{and}
F_{\mu\nu}^{(\phi W)} & \equiv \kappa e g \cos
  \theta_W \allcontpw \, ,
\end{align}
we may write the contributions to the amplitudes from individual
diagrams as
\begin{align}
\diagconthw{a} &= -\lint
N_{\alpha\beta\mu}(l,-l-p,p)D_W^{\alpha\tau}(l)c^W_{\lambda\rho\sigma\tau}(l+q,l)
\times \nonumber\\ 
& \hphantom{=-\lint} \times
D_W^{\sigma\delta}(l+q)N_{\gamma\delta\nu}(l+p,-l-q,-k)D_W^{\gamma\beta}(l+p)
\, \label{eq:diagconthwa},\\
\diagconthw{b} &= -\lint
N_{\alpha\beta\mu}(l-k,-l-q,p)D_W^{\alpha\tau}(l-k) \times \nonumber\\
& \hphantom{=-\lint} \times
N_{\sigma\tau\nu}(l,-l+k,-k)D_W^{\delta\sigma}(l)c^W_{\lambda\rho\gamma\delta}(l+q,l)
D_W^{\beta\gamma}(l+q) \, \label{eq:diagconthwb},\displaybreak[1]\\
\diagconthw{c} &= i \lint N_{\alpha\beta\mu}(l,-l-p,p)
D_W^{\alpha\tau}(l) d_{\lambda\rho\sigma\tau\nu}(l+p,-l,-k)
D_W^{\sigma\beta}(l+p) \, \label{eq:diagconthwc},\displaybreak[0]\\
\diagconthw{d} &= i \lint d_{\lambda\rho\alpha\beta\mu}(l,-l-k,p)
D_W^{\tau\alpha}(l) N_{\sigma\tau\nu}(l+k,-l,-k)
D_W^{\beta\sigma}(l+k) \, \label{eq:diagconthwd},\displaybreak[0]\\
\diagconthw{e} &=
-D^{\gamma\delta}(p)c_{\lambda\rho\delta\nu}(p,k) \times \nonumber \\
&\hphantom{=} \times \lint
N_{\alpha\beta\mu}(l,-l-p,p)D_W^{\tau\alpha}(l)
N_{\sigma\tau\gamma}(l+p,-l,-p) D_W^{\beta\sigma}(l+p) \, \label{eq:diagconthwe},\displaybreak[0]\\
\diagconthw{f} &= -c^Z_{\lambda\rho\mu\alpha}(p,k)D_Z^{\alpha\beta}(k)
  \times \nonumber\\
&\hphantom{=} \times \lint N_{\gamma\delta\beta}(l,-l-k,k) D_W^{\tau\gamma}(l)
  N_{\sigma\tau\nu}(l+k,-l,-k) D_W^{\delta\sigma}(l+k) \, \label{eq:diagconthwf},\displaybreak[0]\\
\diagconthw{g} &= -iR_{\alpha\beta\mu\nu}\lint D_W^{\tau\alpha}(l)
c^W_{\lambda\rho\sigma\tau}(l+q,l) D_W^{\beta\sigma}(l+q) \, \label{eq:diagconthwg},\displaybreak[1]\\
\diagconthw{h} &= -R_{\lambda\rho\alpha\beta\mu\nu}\lint
D_W^{\alpha\beta}(l) \, \label{eq:diagconthwh},\displaybreak[0]\\
\diagconthw{i} &= -ic_{\lambda\rho\alpha\nu}(p,k) D^{\alpha\beta}(p)
R_{\sigma\tau\mu\beta} \lint D_W^{\sigma\tau}(l) \, \label{eq:diagconthwi},\\
\diagconthw{j} &= -ic^Z_{\lambda\rho\mu\alpha}(p,k)
D_Z^{\alpha\beta}(k) R_{\sigma\tau\beta\nu} \lint D_W^{\sigma\tau}(l)
\, \label{eq:diagconthwj},
\intertext{and}
\diagcontpw{a} &= -2\omega M_W^2 \eta_{\sigma\tau} \lint
N_{\alpha\beta\mu}(l,-l-p,p) D_W^{\alpha\tau}(l)
D_W^{\delta\sigma}(l+q) \times \nonumber\\
& \hphantom{=-2\omega M_W^2 \eta_{\sigma\tau} \lint} \times N_{\gamma\delta\nu}(l+p,-l-q,-k)
D_W^{\beta\gamma}(l+p) \, \label{eq:diagcontpwa},\\
\diagcontpw{b} &= -2\omega M_W^2 \eta_{\gamma\delta} \lint
N_{\alpha\beta\mu}(l-k,-l-q,p) D_W^{\alpha\tau}(l-k) \times \nonumber\\
& \hphantom{= -2\omega M_W^2 \eta_{\gamma\delta} \lint} \times
N_{\sigma\tau\nu}(l,-l+k,-k) D_W^{\delta\sigma}(l)
D_W^{\beta\gamma}(l+q)\, \label{eq:diagcontpwb},\\
\diagcontpw{f} &= -2\omega M_Z^2 \eta_{\mu\alpha}D_Z^{\alpha\beta}(k)
\times \nonumber\\
&\hphantom{=} \times \lint N_{\gamma\delta\beta}(l,-l-k,k) D_W^{\tau\gamma}(l)
N_{\sigma\tau\nu}(l+k,-l,-k) D_W^{\delta\sigma}(l+k) \, \label{eq:diagcontpwf},\\
\diagcontpw{g} &= -2i\omega M_W^2 \eta_{\sigma\tau}
R_{\alpha\beta\mu\nu} \lint D_W^{\tau\alpha}(l) D_W^{\beta\sigma}(l+q)
\, \label{eq:diagcontpwg},\\
\diagcontpw{j} &= -2i \omega M_Z^2 \eta_{\mu\alpha}
R_{\sigma\tau\beta\nu} D_Z^{\alpha\beta}(k) \lint D_W^{\tau\sigma}(l)
\, \label{eq:diagcontpwj};
\end{align}
we have omitted a number of lines corresponding to diagrams involving
$\ptiln$ production that evaluate to zero.

\section{Counterterm diagrams}\label{sec:countertermdiagrams}

Because we are working at one loop using bare parameters, we must
consider corrections that arise from a Standard Model Z-$\gamma$
mixing counterterm. It will turn out that such terms give a zero
contribution (and this is why we work with bare parameters), and it is
sufficient just to consider the general form that such terms take in
order to demonstrate this.

%%%%%%%%%%%%%%%%%%%%%%%%%%%%%%%%%%%%%%%%%%%%
% htilde counterterm diagrams
\begin{figure}
%\FIGURE{
\begin{center}
\subfloat[][]{\twoendvertexcounterterm{zboson}{photon}{dbl_wiggly}}
\subfloat[][]{\threeendcounterterm{zboson}{photon}{dbl_wiggly}}
\subfloat[][]{\twoendcountertermvertex{zboson}{photon}{dbl_wiggly}}
\caption{Diagrams for the process $Z \to \gamma + \htiln$ involving
  counterterm vertices.}\label{fig:htilcounterterms}
\end{center}
%}
\end{figure}
% end htilde counterterm diagrams
%%%%%%%%%%%%%%%%%%%%%%%%%%%%%%%%%%%%%%%%%%%%
Figure~\ref{fig:htilcounterterms} contains the diagrams for the
$Z\to\gamma +\htiln$ process that have a counterterm, and
figure~\ref{fig:ptilcounterterms} contains the diagrams for the
$Z\to\gamma +\ptiln$ process that have a counterterm.
%%%%%%%%%%%%%%%%%%%%%%%%%%%%%%%%%%%%%%%%%%%%
% phitilde counterterm diagrams
\begin{figure}
%\FIGURE{
\begin{center}
\subfloat[][]{\twoendvertexcounterterm{zboson}{photon}{double}}
\subfloat[][]{\threeendcounterterm{zboson}{photon}{double}}
\subfloat[][]{\twoendcountertermvertex{zboson}{photon}{double}}
\caption{Diagrams for the process $Z \to \gamma + \ptiln$ involving
  counterterm vertices.}\label{fig:ptilcounterterms}
\end{center}
%}
\end{figure}
% end phitilde counterterm diagrams
%%%%%%%%%%%%%%%%%%%%%%%%%%%%%%%%%%%%%%%%%%%%
We may divide the diagrams into two classes: those containing a
two-point counterterm vertex and those containing a three-point
counterterm vertex. The two-point counterterm vertex is the
Z-$\gamma$ mixing counterterm that occurs in the Standard Model if one
uses the tree-level diagonalization of the electroweak mixing matrix
for one-loop calculations. The three-point counterterm vertex arises
by considering the gravitational perturbation expansion about the
Lagrangian term corresponding to the two-point counterterm vertex.

To derive the Feynman rules for the counterterm vertices in this
regime, we need to consider the Lagrangian Z-$\gamma$ counterterm that
arises from one-loop renormalization in the Standard
Model~\cite{AoHKKM1982}. The relevant terms in the bare
Standard Model Lagrangian may be written as
\begin{equation}
\mathcal{L}_{0\, {\rm Z},\gamma} = -\frac{1}{2} Z_{\mu}
(-\eta^{\mu\nu}\partial^2+\partial^{\mu}\partial^{\nu}) Z_{\nu}-\frac{1}{2} A_{\mu}
(-\eta^{\mu\nu}\partial^2+\partial^{\mu}\partial^{\nu}) A_{\nu} +
\frac{1}{2}M_Z^2 Z_{\mu} \eta^{\mu\nu} Z_{\nu} \, ,
\end{equation}
and applying the renormalization
\begin{align}
Z_{\mu} &\to Z_{ZZ}^{1/2}Z_{\mu}+Z_{ZA}^{1/2}A_{\mu} \, ,\\
A_{\mu} &\to Z_{AZ}^{1/2}Z_{\mu}+Z_{AA}^{1/2}A_{\mu} \, ,\\
M_Z^2 &\to  M_Z^2 + \delta M_Z^2 \, ,\\
\intertext{we obtain a mixing counterterm in the Lagrangian, which may be written}
\mathcal{L}_{\rm{Z}\gamma} &= \fourzs \left[
  \partial_{\nu}Z^{\mu}\partial_{\mu}A^{\nu} -
  \eta_{\mu\nu}\partial^{\alpha}Z^{\mu}\partial_{\alpha}A^{\nu}
  \right] + \nonumber\\
 &\hphantom{=} +\msdms \twozsnobracket \eta_{\mu\nu}Z^{\mu}A^{\nu} \, .
\end{align}
We may read off from this that the Feynman rule for the two-point
Z-$\gamma$ counterterm vertex with
momentum $k$ passing through is~\cite{AoHKKM1982}
\begin{align}
&\left( \eta_{\mu\nu} - \frac{k_{\mu}k_{\nu}}{k^2} \right)
\left[ \msdms
 \twozs - k^2 \fourzs \right] + \nonumber\\
& + \frac{k_{\mu}k_{\nu}}{k^2} \left[ \msdms
 \twozs \right] .
\end{align}
We write
this contribution in the form
\begin{equation}
A(k)\eta_{\mu\nu} + B k_{\mu}k_{\nu} \, ,
\end{equation}
where
\begin{align}
A(k) &= \msdms \twozsnobracket - k^2 \fourzs \label{eq:countertermadefine}\\
\intertext{and}
B &= \fourzsnobracket \, \label{eq:countertermbdefine}.
\end{align}
For the three-point vertices, we must consider the gravitational
coupling expansion of the metric. We may write the $O(\kappa)$ term of
the expanded Lagrangian as~\cite{HaLZ1998}
\begin{align}
\mathcal{L}_{\kappa} =& -\kappa \sum_{\vec{n}} \int d^4 x \left(
 \htilup T_{\lambda\rho} + \omega
 \ptiln T^{\lambda}_{\phantom{\lambda} \lambda} \right) ,\\
\intertext{where}
T_{\lambda\rho} =& \left. \left( -\eta_{\lambda\rho} \mathcal{L} + 2 \frac{\delta
 \mathcal{L}}{\delta g^{\lambda\rho}} \right) \right|_{g=\eta} \, ,
\end{align}
and we have replaced Minkowski metric terms $\eta_{\mu\nu}$ in the
Lagrangian with the perturbed metric $g_{\mu\nu}$. We find that the Z-$\gamma$
mixing terms in the energy-momentum tensor are
\begin{align}
T^{({\rm Z}\gamma)}_{\lambda\rho} =& \fourzs \times \nonumber\\
  & \times \left[
  \eta_{\mu\nu}\eta_{\lambda\rho}\partial^{\alpha}Z^{\mu}\partial_{\alpha}A^{\nu}
  - \eta_{\lambda\rho}\partial_{\nu}Z^{\mu}\partial_{\mu}A^{\nu} -
  \eta_{\mu\lambda}\eta_{\nu\rho}\partial^{\alpha}Z^{\mu}\partial_{\alpha}A^{\nu} -
  \right.\nonumber\\
 & \hphantom{\times [} -
  \eta_{\mu\rho}\eta_{\nu\lambda}\partial^{\alpha}Z^{\mu}\partial_{\alpha}A^{\nu}
  - \eta_{\mu\nu}\partial_{\lambda}Z^{\mu}\partial_{\rho}A^{\nu} -
  \eta_{\mu\nu}\partial_{\rho}Z^{\mu}\partial_{\lambda}A^{\nu} +
  \nonumber\\
 & \hphantom{\times [} \left. +
  \eta_{\mu\lambda}\partial_{\nu}Z^{\mu}\partial_{\rho}A^{\nu} +
  \eta_{\mu\rho}\partial_{\nu}Z^{\mu}\partial_{\lambda}A^{\nu} +
  \eta_{\rho\nu}\partial_{\lambda}Z^{\mu}\partial_{\mu}A^{\nu} +
  \eta_{\lambda\nu}\partial_{\rho}Z^{\mu}\partial_{\mu}A^{\nu}\right]
  + \nonumber\\
 & + \msdms \twozsnobracket \left[
  \eta_{\mu\lambda}\eta_{\rho\nu}+\eta_{\mu\rho}\eta_{\lambda\nu}-\eta_{\mu\nu}\eta_{\lambda\rho}
  \right] Z^{\mu}A^{\nu} \, ,\\
\intertext{with trace}
T^{\lambda}_{\phantom{\lambda}\lambda} =& -2 \msdms \twozsnobracket
  \eta_{\mu\nu} Z^{\mu} A^{\nu} \, .
\end{align}
This yields a Feynman rule for the vertex
$Z^{\mu}(p)A^{\nu}(k)\htilup(q)$ of
\begin{equation}
-\kappa \fourzs c_{\lambda\rho\mu\nu}(p,k) - \kappa \msdms
 \twozsnobracket C_{\mu\nu\lambda\rho} \, ,
\end{equation}
where $C_{\mu\nu\lambda\rho}$ and $c_{\lambda\rho\mu\nu}(p,k)$ are
defined in equations~\eqref{eq:bigcdefine}
and~\eqref{eq:smallcdefine} respectively,
and a Feynman rule for the vertex $Z^{\mu}(p)A^{\nu}(k)\ptiln(q)$ of
\begin{equation}
2\omega \kappa \msdms \twozsnobracket \eta_{\mu\nu} \, .
\end{equation}
We note for future reference that the form of the Z-$\gamma$-$\htiln$
vertex is such that there is no term in which there are four momenta
all carrying Lorentz indices (this will be important in showing that
the counterterms give no contribution to the amplitude).

\setlength{\unitlength}{1cm}
%%%%%%%%%%%%%%%%%%%%%%%%%%%%%%%%%%%%%%
% Extra vertex Feynman rules for the counterterm diagrams
\begin{figure}
%\FIGURE{
\begin{center}
\subfloat{%
\begin{tabular}{c}
\fmfframevertex{%
\begin{fmfgraph*}(2,2)
\fmfsurround{v1,v2}
\fmf{photon_arrw,label=$k$,l.side=left}{i1,v1}
\fmfplab{Z_{\alpha}}{v2}
\fmf{zboson_arrw,label=$k$,l.side=left}{v2,i1}
\fmfplab{A_{\beta}}{v1}
\fmfpen{thick}
\fmfv{decor.shape=cross}{i1}
\fmfpen{thin}
\end{fmfgraph*}}\\
$A(k)\eta_{\alpha\beta} + Bk_{\alpha}k_{\beta}$
\end{tabular}
}
\subfloat{%
\begin{tabular}{c}
\fmfframevertex{%
\begin{fmfgraph*}(2,2)
\fmfsurround{v1,v2,v3}
\fmf{photon_arrw,label=$k$,l.side=left}{i1,v3}
\fmfplab{A_{\nu}}{v3}
\fmf{double}{v2,i1}
\fmfplab{\ptiln}{v2}
\fmf{zboson_arrw,label=$p$,l.side=right}{v1,i1}
\fmfplab{Z_{\mu}}{v1}
\fmfpen{thick}
\fmfv{decor.shape=cross}{i1}
\fmfpen{thin}
\end{fmfgraph*}}\\
$2\omega \kappa \msdms \twozsnobracket \eta_{\mu\nu}$
\end{tabular}
}
\subfloat{%
\begin{tabular}{c}
\fmfframevertex{%
\begin{fmfgraph*}(2,2)
\fmfsurround{v1,v2,v3}
\fmf{photon_arrw,label=$k$,l.side=left}{i1,v3}
\fmfplab{A_{\nu}}{v3}
\fmf{dbl_wiggly}{v2,i1}
\fmfplab{\htil}{v2}
\fmf{zboson_arrw,label=$p$,l.side=right}{v1,i1}
\fmfplab{Z_{\mu}}{v1}
\fmfpen{thick}
\fmfv{decor.shape=cross}{i1}
\fmfpen{thin}
\end{fmfgraph*}}\\
$-\kappa \fourzs c_{\lambda\rho\mu\nu}(p,k) - \kappa \msdms
 \twozsnobracket C_{\mu\nu\lambda\rho}$
\end{tabular}
}
\caption{Extra Feynman rules for counterterm vertices. Abbreviations
  are defined in the main text.}
\label{fig:countertermfeynman}
\end{center}
%}
\end{figure}
% end extra vertex Feynman rules for the counterterm diagrams
%%%%%%%%%%%%%%%%%%%%%%%%%%%%%%%%%%%%%%%%%%%
\setlength{\unitlength}{1.8cm}
The additional Feynman rules are summarized in figure~\ref{fig:countertermfeynman}.

With these Feynman rules, and writing $\allconthc$ and
$\allcontpc$ to denote the contributions from
the counterterm diagrams to the off-shell amplitudes
$F_{\lambda\rho\mu\nu}^{(h)}$ and $F_{\mu\nu}^{(\phi )}$,
respectively, we may write the contributions from
individual diagrams to the amplitudes as
\begin{align}
\diagconthc{a} &= \kappa
c^Z_{\lambda\rho\mu\alpha}(p,k)D_Z^{\alpha\beta}(k) \left[
  A(k)\eta_{\beta\nu} + B k_{\beta} k_{\nu} \right] \label{eq:diagconthca},\\
\diagconthc{b} &= i\kappa \left[ \fourzs c_{\lambda\rho\mu\nu}(p,k) + \msdms
 \twozsnobracket C_{\mu\nu\lambda\rho}\right] \label{eq:diagconthcb},\\
\diagconthc{c} &= \kappa c_{\lambda\rho\beta\nu}(p,k)
D^{\alpha\beta}(p) \left[A(p)\eta_{\mu\alpha}+B p_{\mu} p_{\alpha}
  \right] \label{eq:diagconthcc},\\
\intertext{and}
\diagcontpc{a} &= 2\omega\kappa \eta_{\mu\alpha} M_Z^2
D_Z^{\alpha\beta}(k) \left[ A(k)\eta_{\beta\nu} + B k_{\beta} k_{\nu}
  \right] \label{eq:diagcontpca},\\
\diagcontpc{b} &= -2i\omega\kappa \msdms \twozsnobracket \eta_{\mu\nu}
\, \label{eq:diagcontpcb},\\
\diagcontpc{c} &= 0 \, \label{eq:diagcontpcc}.
\end{align}

\section{General forms of the amplitudes}\label{sec:generalform}

We now derive general forms that must be taken by the amplitudes we
are calculating; these forms allow simplification of the
calculation. The argument for the decay involving a spin-2 graviton
excitation is that of Nieves and Pal~\cite{NiP2005}; the argument for
the decay involving a spin-0 graviton excitation is essentially the
first part of the argument for the spin-2 case, and is given first as
it is more straightforward.

The arguments rely upon Ward-Takahashi identities that are
consequences of electromagnetic and gravitational gauge
invariance. This gauge invariance %can be seen relatively easily 
is shown in the next section
for
each of the sets of fermion loop diagrams, W~loop diagrams and
counterterm diagrams separately%, using the arguments contained in
%Sections 3.3, 3.4, 4.3 and 4.4 of reference~\cite{NiP2005}, and by
%direct calculation in the case of the counterterm diagrams%
.

\subsection{Decay into spin-0 excitation and photon}

Electromagnetic gauge invariance implies conservation of the
electromagnetic current, which (transforming to momentum space) gives
\begin{equation}
k^{\nu}F^{(\phi)}_{\mu\nu}(q,k) = 0 \,\label{eq:emtransphi} .
\end{equation}
We may expand $F^{(\phi)}_{\mu\nu}$ about $k=0$, writing
\begin{equation}
F^{(\phi)}_{\mu\nu} =
\mathcal{T}^0_{\mu\nu}+k^{\alpha}\mathcal{T}^1_{\mu\nu\alpha} \, ,
\end{equation}
with $\mathcal{T}^0_{\mu\nu}$ independent of
$k$. Equation~\eqref{eq:emtransphi} then implies that
\begin{align}
k^{\nu}\mathcal{T}^0_{\mu\nu} &= 0 \, ,\\
k^{\nu}k^{\alpha}\mathcal{T}^1_{\mu\nu\alpha} &= 0 \, ,
\end{align}
and since this is true for all $k$ with $|k_0|\leq M_Z$ in
the centre of mass frame and $k^2=0$, we
can deduce that
\begin{align}
\mathcal{T}^0_{\mu\nu} &=  0 \\
\intertext{and}
\mathcal{T}^1_{\mu\nu\alpha} &=  -\mathcal{T}^1_{\mu\alpha\nu} \,
.
\end{align}
This means we may write the on-shell amplitude \eqref{eq:pmatelement} as
\begin{align}
\matel^{(\phi )}(q,k) &= \polz f^{\nu\alpha *}t_{\mu\nu\alpha} \, ,\\
\intertext{where}
f_{\nu\alpha} &\equiv k_{\nu}\varepsilon_{\alpha} -
k_{\alpha}\varepsilon_{\nu}\label{eq:genformemf}\\
\intertext{and}
t_{\mu\nu\alpha} &= -t_{\mu\alpha\nu} \, .
\end{align}
Recalling equations~\eqref{eq:photonpol} and~\eqref{eq:photonmomsq}
($\polp k_{\nu}=0$ and $k^2=0$), along with the momentum conservation
relation $p=k+q$, and considering terms that could be contained in
$t_{\mu\nu\alpha}$, we see that the only terms contributing to the
amplitude are ones not involving a Levi-Civita tensor, of the form
\begin{equation}
\mbox{scalar} \times \polz (k^{\nu}\varepsilon^{\alpha *} -
k^{\alpha}\varepsilon^{\nu *})(\eta_{\mu\nu}q_{\alpha} -
\eta_{\mu\alpha}q_{\nu}) \, ,
\end{equation}
and ones involving a Levi-Civita tensor, of the form
\begin{equation}
\mbox{scalar} \times \polz
\epsilon^{\nu\alpha\gamma\delta}(k_{\gamma}\varepsilon_{\delta *} -
k_{\delta}\varepsilon_{\gamma *})(\eta_{\mu\nu}q_{\alpha} -
\eta_{\mu\alpha}q_{\nu}) \, .
\end{equation}
Relabelling indices so that the polarisation tensors may be written as
common coefficients of the overall amplitude, we may rewrite the above as
\begin{equation}
F^{(\phi)}_{\mu\nu} = (k_{\mu}q_{\nu}-k\cdot q
\eta_{\mu\nu})F^{(\phi)} +
(\epsilon_{\mu\nu\alpha\beta}q^{\alpha}k^{\beta})F_1^{(\phi)}\,\label{eq:spinzerogeneralform} ,
\end{equation}
where $F^{(\phi)}$ and $F_1^{(\phi)}$ are Lorentz scalars.

\subsection{Decay into spin-2 excitation and photon}

The argument in the case of decay involving a spin-2 graviton
excitation may be viewed as an extension of the case involving a
spin-0 graviton excitation. It is given in full in reference~\cite{NiP2005} for
the case of a massless graviton, and
is sketched here.

Similarly to the spin-0 case, we may write
\begin{equation}
F_{\lambda\rho\mu\nu} = \mathcal{T}^0_{\lambda\rho\mu\nu} +
k^{\alpha}\mathcal{T}^1_{\lambda\rho\mu\nu\alpha} \, ,
\end{equation} 
and we may use the condition $k^{\nu}F_{\lambda\rho\mu\nu}$ to deduce
that we may write the on-shell amplitude \eqref{eq:hmatelement} in the form
\begin{equation}
\matel^{(h)}(q,k) = \polg \polz f^{\nu\alpha *} t_{\lambda\rho\mu\nu\alpha} \, ,
\end{equation}
where $f_{\nu\alpha}$ is as defined in
equation~\eqref{eq:genformemf}.

We may also use gravitational gauge
invariance to derive a manner of writing the amplitude, in a similar
fashion to the way we have used electromagnetic invariance (although
the details are complicated by the extra Lorentz index in the
gravitational case).

Writing
\begin{equation}
%j_{\lambda\rho} = \polp \polz F_{\lambda\rho\mu\nu} \, ,
\matel^{(h)} = \polg j_{\lambda\rho}(q,k) \, ,
\end{equation}
and expanding
\begin{equation}
j_{\lambda\rho} = j^0_{\lambda\rho} +
q^{\sigma}j^1_{\lambda\rho\sigma} +
q^{\tau}q^{\sigma}j^2_{\lambda\rho\sigma\tau} \, \label{eq:gravpolexpansion} ,
\end{equation}
where $j^0_{\lambda\rho}$ and $j^1_{\lambda\rho\sigma}$ are
independent of $q$, we may use the conditions
\begin{equation}
q^{\lambda}j_{\lambda\rho}=0, \qquad q^{\rho}j_{\lambda\rho}=0
\end{equation}
(following from conservation of the energy-momentum tensor) to deduce
\begin{align}
j^0_{\lambda\rho} &= 0 \, ,\\
j^1_{\lambda\rho\sigma} &= 0 \, .
\end{align}
In addition, $j^2_{\lambda\rho\sigma\tau}$ has a number of symmetries
arising from its definition (namely $\lambda \leftrightarrow \rho$
symmetry, $\sigma \leftrightarrow \tau$ symmetry and antisymmetry
under interchange of either of $(\lambda,\rho)$ with either of
$(\sigma,\tau)$). We also note that no term in the expansion of
$j^2_{\lambda\rho\sigma\tau}$ contains a $q_{\lambda}$ or $q_{\rho}$
term, because such terms vanish on contraction with the gravitational
polarisation tensor (equation~\eqref{eq:gravpol}). Further, we need
not consider terms in $q_{\sigma}$ or $q_{\tau}$, since $q^2=\kkm^2$,
and so such terms reduce to lower order terms in equation~\eqref{eq:gravpolexpansion}.
Combining these properties with the expression for the form of the amplitude obtained
from consideration of the electromagnetic gauge invariance yields a
general form of the amplitude that may be written as
\begin{align}
F^{(h)}_{\lambda\rho\mu\nu} =& \left\{ (k_{\lambda}q_{\nu} -
k\cdot q \eta_{\nu\lambda}) (k_{\rho}q_{\mu} - k \cdot q
\eta_{\mu\rho} )F^{(h)}+
\epsilon_{\lambda\nu\alpha\beta}q^{\alpha}k^{\beta} (k_{\rho}q_{\mu} -
k\cdot q \eta_{\mu\rho}) F^{(h)}_1+\right. \nonumber\\
& \phantom{\{}\left. +   (k_{\lambda}q_{\nu} - k\cdot q
\eta_{\nu\lambda}) \epsilon_{\rho\mu\alpha\beta}q^{\alpha}k^{\beta}F^{(h)}_2
\right\} + (\lambda \leftrightarrow \rho) \, \label{eq:generalformspintwo}.
\end{align}

\section{Transversality conditions}\label{sec:transversality}

In order to apply to any set of diagrams the arguments presented in
section~\ref{sec:generalform} regarding the general forms of the
amplitudes, we must establish that if we take the contribution given
by the set of diagrams to the amplitude and contract it with the
photon momentum, this yields zero. (In the case of the decays
involving a spin-2 graviton excitation, we also need to establish that
contraction with the graviton momentum yields zero.)

\subsection{Electromagnetic transversality: spin-2 case}

We begin by considering the case of decays involving production of a
spin-2 graviton excitation. In the spin-2 case, the contributions to
the amplitude of the diagrams involving a fermion loop and of the
diagrams involving a W~boson loop are essentially the same as the contributions
given in~\cite{NiP2005}. We repeat here for completeness an outline of
the derivation of the transversality results in those cases. The
counterterm diagrams are not considered explicitly in~\cite{NiP2005}, and we present more
fully the argument for these diagrams. 

\subsubsection{Fermion loop diagrams}

Using the identity\footnote{The identities differ from those given
  in~\cite{NiP2005} by factors of $i$ owing to the different
  propagator definitions.}
\begin{equation}
k^{\nu}S(l)\gamma_{\nu}S(l+k)=iS(l)-iS(l+k)\label{eq:fermionpropagatorgammaid}
\end{equation}
with equation~\eqref{eq:bigpiintegraldefine} implies that
\begin{equation}
k^{\nu} \Pi_{\mu\nu}(k)=0 \, \label{eq:bigpiemtrans},
\end{equation}
from which, using equations~\eqref{eq:diagconthfd} and~\eqref{eq:diagconthff}, it follows that
\begin{equation}
k^{\nu}\diagconthf{d} = 0, \qquad k^{\nu}\diagconthf{f} = 0 \, .
\end{equation}
Using the definition of equation~\eqref{eq:smallcdefine}, we also have
\begin{equation}
k^{\nu}c_{\lambda\rho\beta\nu}(p,k)=0 \label{eq:smallcpmomcontract} \, ,
\end{equation}
from which it follows that
\begin{equation}
k^{\nu}\diagconthf{e}=0 \, .
\end{equation}
We can show that
\begin{equation}
k^{\nu}\diagconthf{a}+k^{\nu}\diagconthf{b}+k^{\nu}\diagconthf{c}=0 \, ,
\end{equation}
by using equation~\eqref{eq:fermionpropagatorgammaid} on the sum
$k^{\nu}\diagconthf{a}+k^{\nu}\diagconthf{b}$ and writing the
resultant term so that the only propagator terms are $S(l)$ and
$S(l+p)$, and then using the Ward-Takahashi identity
\begin{equation}
V_{\lambda\rho}(l+q,l)- V_{\lambda\rho}(l+p,l+k) =
a_{\lambda\rho\alpha\beta}\gamma^{\alpha}k^{\beta} \, .
\end{equation}
This establishes that
\begin{equation}
k^{\nu}\allconthf=0 \, .
\end{equation}

\subsubsection{W loop diagrams}

Electromagnetic transversality of the W~loop diagrams contributing to
the spin-2 excitation producing decay may be demonstrated by using a
number of Ward-Takahashi identities. The identities
\begin{align}
r^{\gamma}D_W^{\alpha\sigma}(l+r)N_{\alpha\beta\gamma}(l+r,-l,-r)D_W^{\beta\tau}(l)
&= iD_W^{\sigma\tau}(l) - iD_W^{\sigma\tau}(l+r) \, \label{eq:wdiagramswardidentity},\\
c^W_{\lambda\rho\alpha\beta}(l+s,l)-c^W_{\lambda\rho\alpha\beta}(l+r+s,l+r)
&= r^{\gamma}d_{\lambda\rho\alpha\beta\gamma}(l+r+s,-l,-r) \,
\label{eq:wpropagatorid} ,\\
\intertext{and}
N_{\alpha,\beta,\gamma}(l-r,-l-s,r+s)-N_{\alpha,\beta,\gamma}(l,-l-r-s,r+s)
&= r^{\delta}R_{\alpha\beta\gamma\delta} \label{eq:wvertexfactorsid}\, ,
\end{align}
together with equation~\eqref{eq:smallcpmomcontract}, can be used with
equations~\eqref{eq:diagconthwa} to~\eqref{eq:diagconthwj} to
deduce that
\begin{align}
k^{\nu}\diagconthw{a}+k^{\nu}\diagconthw{b}+k^{\nu}\diagconthw{c}+k^{\nu}\diagconthw{g}
&= 0 \, ,\\
k^{\nu}\diagconthw{d}+k^{\nu}\diagconthw{h} &= 0 \, ,\\
k^{\nu}\diagconthw{e} &= 0 \, ,\\
k^{\nu}\diagconthw{f}+k^{\nu}\diagconthw{j} &= 0 \, ,\\
\intertext{and}
k^{\nu}\diagconthw{i} &= 0 \, ,\\
\intertext{and therefore that}
k^{\nu}\allconthw &= 0 \, .
\end{align}

\subsubsection{Counterterm diagrams}

Using equation~\eqref{eq:smallcpmomcontract} with the definition of equation~\eqref{eq:diagconthcc} implies that
\begin{equation}
k^{\nu}\diagconthc{c} = 0 \, .
\end{equation}
Equation~\eqref{eq:smallcpmomcontract} also allows us to simplify the
forms of $k^{\nu}\diagconthc{a}$ and $k^{\nu}\diagconthc{b}$. If we
expand the definitions of $c^Z_{\lambda\rho\mu\alpha}(p,k)$ and
$D_Z^{\alpha\beta}(k)$ from equations~\eqref{eq:smallczdefine}
and~\eqref{eq:zpropagatordefine} respectively, this yields
\begin{equation}
k^{\nu}\diagconthc{a} = \kappa A(k) (k_{\mu}\eta_{\lambda\rho} -
k_{\lambda}\eta_{\mu\rho} - k_{\rho}\eta_{\mu\lambda}) =
-k^{\nu}\diagconthc{b} \, ,
\end{equation}
from which we deduce that
\begin{equation}
k^{\nu}\diagconthc{a}+k^{\nu}\diagconthc{b}+k^{\nu}\diagconthc{c} = 0
\, .
\end{equation}

\subsection{Gravitational transversality: spin-2 case}
\subsubsection{Fermion loop diagrams}
Gravitational transversality of the fermion loop diagrams contributing
to the spin-2 graviton excitation production process may be seen
through use of the identities
\begin{align}
S(l)q^{\lambda}V_{\lambda\rho}(l+q,l)S(l+q) &=
i(l_{\rho}+q_{\rho})S(l+q) - il_{\rho}S(l) + \nonumber\\
&\phantom{=} +
\frac{i}{8}\left(\gamma_{\rho}\slashed{q}-\slashed{q}\gamma_{\rho}\right)S(l+q)
 -
 \frac{i}{8}S(l)\left(\gamma_{\rho}\slashed{q}-\slashed{q}\gamma_{\rho}\right) \\
\intertext{and}
\gamma_{\alpha}\gamma_{\beta}\gamma_{\rho}+\gamma_{\rho}\gamma_{\beta}\gamma_{\alpha}
&= 2\left( \eta_{\alpha\beta}\gamma_{\rho}+\eta_{\beta\rho}\gamma_{\alpha}-\eta_{\alpha\rho}\gamma_{\beta}  \right),
\end{align}
together with the momentum conditions and with changes in the
integration variable in some of the loop momenta integrations. Fuller
details may be found in reference~\cite{NiP2005}.

\subsubsection{W loop diagrams}
If we use cyclic identity
\begin{equation}
R_{\alpha\beta\gamma\delta}+R_{\beta\gamma\alpha\delta}+R_{\gamma\alpha\beta\delta}=0
\, ,
\end{equation}
together with the identities
\begin{align}
N_{\alpha\beta\gamma}(l,-l-s,s)-N_{\alpha\beta\gamma}(l,-l-s+r,s-r) &=
r^{\delta}R_{\beta\gamma\alpha\delta} \\
\intertext{and}
N_{\alpha\beta\gamma}(l,-l-s,s)-N_{\alpha\beta\gamma}(l-r,-l-s,r+s) &=
r^{\delta}R_{\gamma\alpha\beta\delta} \, ,
\end{align}
the identity of equation~\eqref{eq:wpropagatorid} and the identity
\begin{align}
q^{\lambda} D_W^{\alpha\sigma}(l+q) c^W_{\lambda\rho\alpha\beta}
D_W^{\beta\tau} &= i(l_{\rho}+q_{\rho})D_W^{\sigma\tau}(l+q) -
il_{\rho}D_W^{\sigma\tau}(l) - \nonumber\\
 & \phantom{=} -
i\eta_{\rho\gamma}\left(q^{\sigma}D_W^{\tau\gamma}(l) +
q^{\tau}D_W^{\sigma\gamma}(l+q)\right) , 
\end{align}
then it is possible to derive the result
\begin{equation}
q^{\lambda} \allconthw = 0 \, .
\end{equation}
More details of the manipulations involved may be found in reference~\cite{NiP2005}.

\subsubsection{Counterterm diagrams}

Gravitational transversality of the counterterm diagrams may be shown
simply by contracting each of the diagrams with $q^{\lambda}$ and
summing the three resultant expressions, if one uses the on-shell
momentum conditions for the Z~boson and the photon
(equations~\eqref{eq:zmomsq} and~\eqref{eq:photonmomsq})~\cite{Ve2000}.

\subsection{Electromagnetic transversality: spin-0 case}

We now turn to demonstrating electromagnetic transversality of the
separate classes of diagrams in the case of decays involving spin-0
graviton excitation production. This is more straightforward than the
spin-2 case, because there are fewer non-zero diagrams to
consider. (In addition, we only need to demonstrate electromagnetic
transversality to establish that the contributions from the classes of
diagrams are of the general form given in
equation~\eqref{eq:spinzerogeneralform}: there is no corresponding
gravitational transversality condition to be satisfied). We
use some of the identities from the spin-2 case.

\subsubsection{Fermion loop diagrams}
Using equation~\eqref{eq:bigpiemtrans} with
equations~\eqref{eq:diagcontpfd} and~\eqref{eq:diagcontpff} implies that
\begin{equation}
k^{\nu}\diagcontpf{d} = 0\, , \qquad k^{\nu}\diagcontpf{f} = 0 \, .
\end{equation}
In addition, we can use equation~\eqref{eq:fermionpropagatorgammaid}
with equations~\eqref{eq:diagcontpfa} and~\eqref{eq:diagcontpfb}
to write
\begin{align}
k^{\nu}\diagcontpf{a} &= 2i\omega \linttr \left\{ S(l-k) - S(l)
\right\} \left( \frac{3}{2} \slashed{l} + \frac{3}{4}\slashed{q} -
2m_f \right) S(l+q) \right] \\
\intertext{and}
k^{\nu}\diagcontpf{b} &= 2i\omega \linttr S(l-q) \left(
\frac{3}{2}\slashed{l} - \frac{3}{4}\slashed{q} - 2m_f \right) \left\{
S(l) - S(l+k) \right\} \right] ,
\end{align}
and we may redefine the loop momenta in the integrands to obtain
\begin{align}
k^{\nu}\diagcontpf{a}+k^{\nu}\diagcontpf{b} &= 3i\omega k^{\nu}
\linttr S(l) \gamma_{\nu} S(l+p) \right] \\
 &= -k^{\nu}\diagcontpf{c} \, ,
\end{align}
so that overall
\begin{equation}
k^{\nu}\allcontpf = 0 \, .
\end{equation}

\subsubsection{W loop diagrams}

We begin by taking the expression for $k^{\nu}\diagcontpw{f}$, using
equation~\eqref{eq:diagcontpwf}, and applying
equation~\eqref{eq:wdiagramswardidentity}, obtaining
\begin{equation}
k^{\nu}\diagcontpw{f} = 2iM_Z^2 \eta_{\mu\alpha} D_Z^{\alpha\beta}(k)
\lint
N_{\gamma\delta\beta}(l,-l-k,k)\left[D_W^{\gamma\delta}(l+k)-D_W^{\gamma\delta}(l)\right] .
\end{equation}
We then redefine the loop momentum
integration variable so that the integrand has a common coefficient
$D_W^{\gamma\delta}(l)$, and apply equation~\eqref{eq:wvertexfactorsid},
to obtain
\begin{align}
k^{\nu}\diagcontpw{f} &= 2i\omega M_Z^2
\eta_{\mu\alpha}k^{\nu}R_{\gamma\delta\beta\nu}D_Z^{\alpha\beta}(k) \lint
D_W^{\gamma\delta}(l) \\
&= -k^{\nu}\diagcontpw{j} \, .
\end{align}
We also redefine the loop momenta integration variables in the
expressions for $\diagcontpw{a}$ and $\diagcontpw{b}$ (equations~\eqref{eq:diagcontpwa}
and~\eqref{eq:diagcontpwb}, taking $l \to
l-q$ and $l \to l+k$, respectively), and use
equation~\eqref{eq:wdiagramswardidentity}, to write
\begin{align}
k^{\nu}\diagcontpw{a}+k^{\nu}\diagcontpw{b} &= 2i\omega M_W^2
\eta_{\sigma\tau} \lint N_{\alpha\beta\mu}(l-q,-l-k,p)
D_W^{\alpha\tau}(l-q) \times \nonumber\\
& \phantom{=2i\omega M_W^2\eta_{\sigma\tau} \lint} \times \left[ D_W^{\beta\sigma}(l+k) -
  D_W^{\beta\sigma}(l) \right] + \nonumber\\
& \phantom{=} + 2i\omega M_W^2 \eta_{\gamma\delta} \lint
N_{\alpha\beta\mu}(l,-l-p,p)D_W^{\beta\gamma}(l+p)\times \nonumber\\
& \phantom{=+2i\omega M_W^2 \eta_{\gamma\delta} \lint} \times \left[
  D_W^{\alpha\delta}(l+k) - D_W^{\alpha\delta}(l) \right] \\
&= 2i\omega M_W^2 \eta_{\sigma\tau} \lint D_W^{\alpha\tau}(l) \left[
  N_{\alpha\beta\mu}(l-k,-l+k-p,p)-\right. \nonumber\\
&\phantom{2i\omega M_W^2 \eta_{\sigma\tau} \lint D_W^{\alpha\tau}(l)} \left.- N_{\alpha\beta\mu}(l,-l-p,p)
  \right] D_W^{\beta\sigma}(l+q) \\
&= 2i\omega M_W^2 \eta_{\sigma\tau}k^{\nu}R_{\alpha\beta\mu\nu} \lint
D_W^{\alpha\tau}(l)D_W^{\beta\sigma}(l+q) \label{eq:phiwemtransintermediate}\\
&= -k^{\nu}\diagcontpw{g} \, ,
\end{align}
where to obtain equation~\eqref{eq:phiwemtransintermediate} we have
used equation~\eqref{eq:wvertexfactorsid}.

As the other diagrams evaluate to zero, we are now able to establish
the result
\begin{equation}
k^{\nu} \allcontpw = 0 \, .
\end{equation}

\subsubsection{Counterterm diagrams}

Contracting with $k^{\nu}$ the expressions for $\diagcontpc{a}$ and
$\diagcontpc{b}$ given in equations~\eqref{eq:diagcontpca}
and~\eqref{eq:diagcontpcb} respectively, and recalling the expressions for $A(k)$ and $B$
given in equations~\eqref{eq:countertermadefine}
and~\eqref{eq:countertermbdefine} respectively, along with the
expression for the propagator $D_Z^{\alpha\beta}(k)$ given in
equation~\eqref{eq:zpropagatordefine}, it is straightforward to see
that
\begin{equation}
k^{\nu}\diagcontpc{a}+k^{\nu}\diagcontpc{b}=0 \, .
\end{equation}
Given that $\diagcontpc{c}=0$, we can therefore deduce that
\begin{equation}
k^{\nu}\allcontpc = 0 \, .
\end{equation}

\section{Calculation of the amplitude coefficients}\label{sec:amplitudecalculation}

Having shown that the individual sets of diagrams (ones with fermion
loops, ones with W~boson loops, and ones with counterterms) separately
satisfy
the requisite conditions to give a contribution to the amplitudes
matching the given general forms, we can now calculate these
contributions separately. We begin by showing that the contribution to the
amplitude from the counterterm diagrams is zero in both the decay
involving a spin-2 graviton excitation and the decay involving a
spin-0 graviton excitation. This will mean that the contributions from
the other sets of diagrams must be finite, and so we should not be
surprised when we see a ``miraculous'' cancellation of divergences in
the case of decay to a spin-0 graviton excitation. Indeed, we note
that as we have not made any particular assumptions about the number
of fermions, we know that there cannot be a cancellation of infinities
between the fermion loop diagrams and the W~boson loop diagrams. This
means that the contribution from each of the individual sets of
diagrams must be finite.

\subsection{Counterterm diagrams contribution}

\subsubsection{Spin-2 case}

We note that there are no Levi-Civita tensors in any of the
counterterm-containing diagrams that contribute to the decay amplitude
in the spin-2 graviton excitation case
(equations~\eqref{eq:diagconthca} to~\eqref{eq:diagconthcc}). The diagrams therefore give no
contribution to either $F_1^{(h)}$ or $F_2^{(h)}$, and we need only
consider the contribution to $F^{(h)}$.

To consider the contribution to $F^{(h)}$ we need consider only one of
the terms of which it is a coefficient, and (as in~\cite{NiP2005}) we
choose the term $F^{(h)}k_{\lambda}k_{\rho}q_{\mu}q_{\nu}$ (noting that
this term appears twice in the expression given in
equation~\eqref{eq:generalformspintwo}). Equation~\eqref{eq:zpol}
implies that
\begin{equation}
\polz k_{\mu} = -\polz q_{\mu} \, ,
\end{equation}
so that terms requiring consideration also arise from terms of the
form $k_{\lambda}k_{\rho}k_{\mu}q_{\nu}$.

For diagram~(a), we note that the contribution has an overall
coefficient of $k_{\nu}$, and therefore does not contribute to the
amplitude. For diagram~(b), we note that there is no term containing
four momenta with Lorentz indices, so there is no contribution from
this term either. Similarly for diagram~(c), once we note that the
term with a $B$ coefficient contains a factor of $p_{\mu}$ so provides
no contribution to the amplitude, we can see that there is no
remaining term with four Lorentz index-carrying momenta. The
contribution to $F^{(h)}$ from the counterterm diagrams is therefore
zero.

\subsubsection{Spin-0 case}

We note that, as in the spin-2 case, there are no Levi-Civita tensors in any of the
counterterm-containing diagrams that contribute to the decay amplitude
in the spin-0 graviton excitation case
(equations~\eqref{eq:diagcontpca} to~\eqref{eq:diagcontpcc}). The diagrams therefore give no
contribution to $F_1^{(\phi)}$, and we need only consider the
contribution to $F^{(\phi)}$.

Similarly to the spin-2 case, we need consider only one of the terms
of which $F^{(\phi)}$ is a coefficient, and we choose the term
$k_{\mu}q_{\nu}$. In this case, the term is not
repeated in the expression for the general form of the
amplitude. Again, equation~\eqref{eq:zpol} implies that terms
requiring consideration also arise from terms of the form $q_{\mu}q_{\nu}$.

It is sufficient for the counterterm diagrams to note that no diagram expression depends
upon the momentum $q$ of the spin-0 particle, so that there is no
contribution from the counterterms to the amplitude.

\subsection{Loop diagram contributions to the spin-2 amplitude coefficients}

We shall see that there are no Levi-Civita terms to be considered, so
that the contributions to $F_1^{(h)}$ and $F_2^{(h)}$ are zero. In
considering contributions to $F^{(h)}$, we shall again look at terms
of the form $k_{\lambda}k_{\rho}q_{\mu}q_{\nu}$ (or
$k_{\lambda}k_{\rho}k_{\mu}q_{\nu}$).

\subsubsection{Fermion loop diagrams}

The determination of the contribution from diagrams containing fermion
loops is very similar to the massless case considered in~\cite{NiP2005}.

With respect to the contributions to $F_1^{(h)}$ and $F_2^{(h)}$, we
note that for diagrams (c), (d), (e) and (f), any term with a
Levi-Civita tensor also has a metric tensor symmetric in two of the
indices of the Levi-Civita tensor, so that there is no contribution
from any of these diagrams.

With respect to the contributions to $F^{(h)}$, we note that none of
diagrams (c), (d), (e) and (f) contributes a term of the
relevant form: this is straightforward to see for diagrams (c) and
(d), which only depend upon $p$ and $k$ respectively; it is relatively
easy to see for diagrams (e) and (f) if one uses the definitions of
$c_{\lambda\rho\beta\nu}(p,k)$ and $c^Z_{\lambda\rho\mu\alpha}(p,k)$
given in equations~\eqref{eq:smallcdefine}
and~\eqref{eq:smallczdefine}, respectively.

It is therefore necessary
only to consider contributions from diagrams (a) and (b).
The contributions from diagrams (a) and (b) may be written in the form
\begin{align}
\diagconthf{a} &= i \lint \frac{l_{\rho} \hbox{Tr} \left[
    \tilde{\gamma}_{\mu} (\slashed{l} - \slashed{k} + m_f)
    \gamma_{\nu} (\slashed{l} + m_f) \gamma_{\lambda} (\slashed{l} +
    \slashed{q} + m_f)\right]}{\left[(l-k)^2-m_f^2\right]
    \left[(l+q)^2-m_f^2\right] \left(l^2-m_f^2\right)} \, ,\\
\diagconthf{b} &= i \lint \frac{l_{\rho} \hbox{Tr} \left[
    \tilde{\gamma}_{\mu} (\slashed{l}-\slashed{q}+m_f)\gamma_{\lambda}
    (\slashed{l}+m_f)
    \gamma_{\nu}(\slashed{l}+\slashed{k}+m_f)\right]}{\left[(l+k)^2-m_f^2\right]
    \left[(l-q)^2-m_f^2\right] \left(l^2-m_f^2\right)} \, .
\end{align}
As in reference~\cite{NiP2005} we can use the cyclic property of the trace and
charge conjugation relations
\begin{equation}
C^{-1}\gamma^{\mu}C = -\gamma^{\mu \, T} \, , \qquad
C^{-1}\gamma^{\mu}\gamma^5 C = \left( \gamma^{\mu} \gamma^5 \right)^T
\, ,
\end{equation}
to write the sum of these terms as
\begin{align}
\diagconthf{a}+\diagconthf{b} &= 2iX_f \lint \frac{f_{\lambda\rho\mu\nu}(l)}{\left[(l-k)^2-m_f^2\right]
    \left[(l+q)^2-m_f^2\right] \left(l^2-m_f^2\right)} \, ,\\
\intertext{where}
f_{\lambda\rho\mu\nu}(l) &= l_{\rho} \hbox{Tr} \left[ \gamma_{\mu}
    (\slashed{l}-\slashed{k}+m_f)\gamma_{\nu}(\slashed{l}+m_f)
    \gamma_{\lambda} (\slashed{l}+\slashed{q}+m_f) \right] \, .\\
\intertext{As the trace does not contain a $\gamma_5$ term, the
    contribution to the amplitude will not contain a Levi-Civita
    tensor, and so the contributions to $F_1^{(h)}$ and $F_2^{(h)}$
    are zero. A Feynman parameterization yields}
\diagconthf{a}+\diagconthf{b} &= 4iX_f \lint \int_0^1 dx \int_0^{1-x}
    dy \frac{f_{\lambda\rho\mu\nu}(l+xk-yq)}{\left[ l^2 - m_f^2 +
    y(1-x-y)\kkm^2 + xyM_Z^2 \right]^3} \, ,
\end{align}
and considering only the terms we have previously noted, there are no
divergences to be considered and we can integrate with respect to the
loop momentum, obtaining a contribution to $F^{(h)}$ from each fermion
in the theory of
\begin{equation}
F^{(hf)} = - \frac{\kappa e g}{2 \pi^2 \cos \theta_W} Q_f X_f
J(m_f,\kkm,M_Z) \, ,
\end{equation}
where
\begin{equation}
J(X,Y,Z) = \int_0^1 dx \int_0^{1-x} dy \frac{x^2y(1-x-y)}{X^2 -
  y(1-x-y)Y^2 - xyZ^2} \, \label{eq:hintjdefine}.
\end{equation}

\subsubsection{W~loop diagrams}

The determination of contributions from diagrams containing W~boson
loops is also very similar to the massless case considered
in~\cite{NiP2005}.

There are no Levi-Civita terms in the individual diagram
contributions, so the overall contributions to $F_1^{(h)}$ and
$F_2^{(h)}$ are zero. We need therefore only consider the
$k_{\lambda}k_{\rho}q_{\mu}q_{\nu}$-like terms to obtain a
contribution to the coefficient $F^{(h)}$.

The contributions from diagrams (g) and (h) have no dependence on $k$,
so cannot take the form we are considering, and therefore may be
ignored. Similarly to the fermion case, the forms of
$c_{\lambda\rho\alpha\nu}(p,k)$ and $c^Z_{\lambda\rho\mu\alpha}(p,k)$,
  given in equations~\eqref{eq:smallcdefine}
and~\eqref{eq:smallczdefine}, respectively, are such that the
combinations $\diagconthw{e}+\diagconthw{i}$ and
$\diagconthw{f}+\diagconthw{j}$ contain no relevant contributions to
the coefficient $F^{(h)}$ (see~\cite{NiP2005} for more
details). Expanding the expressions for diagrams (c) and (d), we see that
no terms in the expressions contain a $q_{\nu}$
term, so these diagrams provide no relevant contribution. We are
therefore left with only the contributions from diagrams (a) and
(b). The contributions from these diagrams are equal: this may be seen
by applying a change of variable in one of the integrands in the
expressions for $\diagconthw{a}$ and $\diagconthw{b}$, or perhaps more
straightforwardly by observing that the contributions obtained from
the diagrams are independent of the direction in which the W~boson
loop is traversed.

It may be shown after some manipulation~\cite{NiP2005} that the
contribution from diagram (b) is contained in the equation
\begin{align}
\diagconthw{b} &= 8i k_{\lambda}k_{\rho}q_{\mu}q_{\nu} \left( 6 -
\frac{M_Z^2}{M_W^2} \right) \times \nonumber\\
 & \phantom{=} \times \lint \int_0^1 dx \int_0^{1-x} dy
\frac{x^2y (1-x-y)}{\left[ l^2 - M_W^2 + y(1-x-y)\kkm^2 + xyM_Z^2
    \right]^3} + \nonumber\\
& \phantom{=} + \mbox{non-contributing terms} \, ,
\end{align}
and therefore, after performing the integration over the loop
momentum, that the overall contribution from the W~boson loop
diagrams to the coefficient $F^{(h)}$ is
\begin{equation}
F^{(hW)} = \frac{\kappa e g \cos \theta_W}{4\pi^2} \left( 6 -
\frac{M_Z^2}{M_W^2} \right) J(M_W,\kkm,M_Z) \, ,
\end{equation}
where $J(X,Y,Z)$ is defined in equation~\eqref{eq:hintjdefine}.

\subsection{Overall contribution to the spin-2 amplitude coefficients}

We have determined so far that the coefficients $F_1^{(h)}$ and $F_2^{(h)}$ are
zero. We have determined also that the coefficient $F^{(h)}$ is given by
\begin{align}
F^{(h)} &= \frac{\kappa e g}{4\pi^2 \cos \theta_W} \times \nonumber\\
&\phantom{=} \times \left[ \cos^2
  \theta_W \left( 6 - \frac{1}{\cos^2 \theta_W}\right) J(M_W,\kkm,M_Z)
  - 2 \sum_f Q_f X_f J(m_f,\kkm,M_Z) \right]
. \label{eq:hcoefficientintegralform}
\end{align}

To proceed, we approximate the analytically intractable integrals of
the form $J(X,Y,Z)$. We
note that all the integrals in which we are interested take the form
$J(X,\kkm,M_Z)$, and since the mass of the Kaluza-Klein mode is
constrained by $\kkm \leq M_Z$, we may consider the integral only in
the cases where $0 \leq Y \leq Z$.

We consider the expression for $J(X,Y,Z)$ given in
equation~\eqref{eq:hintjdefine}, and repeated here for convenience:
\begin{equation}
J(X,Y,Z) = \int_0^1 dx \int_0^{1-x} dy \frac{x^2y(1-x-y)}{X^2 -
  y(1-x-y)Y^2 - xyZ^2} \, .
\end{equation}
We note that over the range of
integration, the maximum value of the expression $xy$ is $1/4$, and
the maximum value of the expression $y(1-x-y)$ is also $1/4$. This
allows us to consider two limiting cases for approximation of the
integral: $X/(Y+Z) \gg 1/4$ and $X/(Y+Z) \ll 1/4$. With the constraint
$0 \leq Y \leq Z$, it is sufficient to consider the cases $X/Z \gg
1/2$ and $X/Z \ll 1/4$.

To evaluate the case $X/Z \gg 1/2$, we may approximate the integral by
$J(X,0,0)$, which is relatively straightforward to calculate, and
yields the result
\begin{equation}
J(X,0,0)= \frac{1}{360X^2} \, .\label{eq:hintegralbigmass}
\end{equation}

To evaluate the case $X/Z \ll 1/2$, we may approximate the integral by
$J(0,Y,Z)$. We may change the order of integration so as to perform
the $x$~integral first; this is possible analytically. We obtain an
integral polynomial in $(1-y)$ that we can evaluate
analytically. Evaluating this integral, we
obtain
\begin{equation}
J(0,Y,Z) = \frac{1}{12(Z^2-Y^2)} - \frac{Z^2}{8(Z^2-Y^2)^2} +
\frac{Y^2Z^2}{4(Z^2-Y^2)^3} + \frac{Y^4Z^2}{4(Z^2-Y^2)^4} \log \left(
\frac{Y^2}{Z^2} \right) .
\end{equation}
To proceed, we use the identity
\begin{align}
\frac{B^2}{C^2-B^2} &\equiv \frac{C^2}{C^2-B^2} - 1 \\
\intertext{to eliminate all terms with a $B^2$ numerator, as well as the identity}
\log \left( \frac{B^2}{C^2} \right) &= \log \left( 1-
\frac{C^2-B^2}{C^2} \right) ,
\end{align}
and expand the logarithm term (the expansion is valid for $0 < Y \leq
Z$, and the final answer turns out to be valid for $Y=0$ as well). We
obtain
\begin{equation}
J(0,Y,Z) = -\frac{1}{2} \sum_{j=1}^{\infty} \frac{1}{(j+1)(j+2)(j+3)}
Z^{-2j}(Z^2-Y^2)^{j-1} \, .\label{eq:hintegralsmallmass}
\end{equation}

We now return to our expression for $F^{(h)}$ given in
equation~\eqref{eq:hcoefficientintegralform}. We approximate the
integrals by using the form of equation~\eqref{eq:hintegralbigmass}
for $J(M_W,\kkm,M_Z)$ and $J(m_t,\kkm,M_Z)$, and the form of
equation~\eqref{eq:hintegralsmallmass} for the integrals relating to
fermions other than the top quark. This gives
\begin{align}
F^{(h)} &= \frac{\kappa e g}{4\pi^2 \cos \theta_W M_Z^2} \left[
  \frac{1}{360} \left( 6 - \frac{1}{\cos^2 \theta_W} \right) +
  \right. \nonumber\\
 & \phantom{= \frac{\kappa e g}{4\pi^2 \cos \theta_W M_Z^2}} + \left( 5 -
  \frac{40}{3} \sin^2 \theta_W \right) \left( \sum_{j=0}^{\infty}
  \frac{1}{(j+2)(j+3)(j+4)} \left( 1 - \frac{\kkm^2}{M_Z^2} \right)^j
  \right) - \nonumber\\
&  \phantom{= \frac{\kappa e g}{4\pi^2 \cos \theta_W M_Z^2}} -
  \left. \frac{M_Z^2}{m_t^2}\left( \frac{1}{180} - \frac{2}{135}
  \sin^2 \theta_W \right) \right]
\end{align}
(where we have used that $M_W = M_Z \cos \theta_W$). Using $e=g\sin
\theta_W$ and $\sin^2 \theta_W = 0.23$, along with a numerical value
for $M_Z/m_t$~\cite{Ya2006}, we obtain the approximation
\begin{equation}
F^{(h)} = \frac{\kappa e^2}{4\pi^2 M_Z^2} \left[ 0.030 + 4.6\left(
  \sum_{j=0}^{\infty} \frac{1}{(j+2)(j+3)(j+4)} \left( 1 -
  \frac{\kkm^2}{M_Z^2} \right)^j \right) \right] \label{eq:fh}.
\end{equation}
This approximation agrees with that given in reference~\cite{NiP2005} in the
case $\kkm = 0$.

\subsection{Loop diagram contributions to the spin-0 amplitude
  coefficients}

As in the spin-2 case, we shall see that there are no Levi-Civita
terms to be considered, so that the contributions to $F_1^{(\phi)}$
are zero. In considering contributions to $F^{(\phi)}$, we shall again
look at terms of the form $k_{\mu}q_{\nu}$.

\subsubsection{Fermion loop diagrams}

The contributions from diagrams (d) and (f) are zero, since the
contributions only depend upon $k$, so they are not of a form to
contribute to $F^{(\phi)}$ and symmetry considerations show
that they do not contribute to $F_1^{(\phi)}$. A similar argument
shows that diagram (c), which depends only upon $p$, does not
contribute to either of the coefficients.

As diagram (e) trivially gives a zero contribution, it follows that it
is necessary only to consider contributions from diagrams (a) and
(b).

The methodology employed is very similar to that of the spin-2 case,
developed in~\cite{NiP2005}. We begin by writing out expressions for
the fermion propagators, so that we may write the contributions as
\begin{align}
\diagcontpf{a} &= -2 i \omega \lint \frac{\mbox{Tr} \left[
    \tilde{\gamma}_{\mu} \left(\slashed{l} - \slashed{k} + m_f\right)
    \gamma_{\nu} \left(\slashed{l}+m_f\right)
    \left(\frac{3}{2}\slashed{l}+\frac{3}{4}\slashed{q}-2m_f\right)
    \left(\slashed{l}+\slashed{q}+m_f\right) \right]}{\left[ (l+q)^2 -
    m_f^2 \right] \left[(l-k)^2 -m_f^2 \right] \left[l^2-m_f^2\right]}
    \, ,\\
\diagcontpf{b} &= -2 i \omega \lint \frac{\mbox{Tr} \left[
    \tilde{\gamma}_{\mu} \left(\slashed{l}-\slashed{q}+m_f\right)
    \left(\frac{3}{2}\slashed{l}-\frac{3}{4}\slashed{q}-2m_f\right)
    \left(\slashed{l}+m_f\right) \gamma_{\nu}
    \left(\slashed{l}+\slashed{k}+m_f\right)
    \right]}{\left[(l+k)^2-m_f^2\right] \left[(l-q)^2-m_f^2\right]
    \left[l^2-m_f^2\right]} \, .
\end{align}
For the second integral, we change the integration parameter to $-l$,
and apply the cyclic property of the trace, the trace-reversal
invariance of strings of gamma matrices, the invariance of transposing
matrices inside the trace, in addition to the charge conjugation
properties of the matrices. We are able to obtain a term in $Y_f$ that
cancels with the corresponding term in the first integral (so that
there is no Levi-Civita term and therefore no contribution to
$F_1^{(\phi)}$), and a term in $X_f$ equal to the corresponding term
in the first integral. We are therefore able to write
\begin{align}
\diagcontpf{a}+\diagcontpf{b} &= -4i\omega X_f \lint
\frac{f_{\mu\nu}(l)}{\left[(l-k)^2-m_f^2\right]
  \left[(l+q)^2-m_f^2\right] \left[l^2-m_f^2\right]} \, ,\\
\intertext{where}
f_{\mu\nu}(l) &= \mbox{Tr} \left[ \gamma_{\mu}
  \left(\slashed{l}+\slashed{q}+m_f\right)
  \left(\frac{3}{2}\slashed{l}+\frac{3}{4}\slashed{q}-2m_f\right)
  \left(\slashed{l}+m_f\right) \gamma_{\nu}
  \left(\slashed{l}-\slashed{k}+m_f\right) \right] .
\end{align}
Applying a Feynman parameterization and continuing to $D$ dimensions,
the equation becomes
\begin{align}
\diagcontpf{a}+\diagcontpf{b} &= -8i\omega X_f \md \times
\nonumber\\
&\phantom{=} \times \ldint
\int_0^1 dx \int_0^{1-x} dy \frac{f_{\mu\nu}(l+xk-yq)}{\left[ l^2 -
    m_f^2 + y(1-x-y)\kkm^2 + xyM_Z^2\right]^3} \, .
\end{align}
We are interested in terms that will contribute to the coefficient of
$k_{\mu}q_{\nu}$. We may use the symmetry properties of the
$l$~integral to discard terms that are odd in $l$, and to replace
\begin{align}
l_{\alpha}l_{\beta} &\to \frac{1}{D}\eta_{\alpha\beta} l^2 \,
\label{eq:twolintreplace},\\
l_{\alpha}l_{\beta}l_{\gamma}l_{\delta} &\to \frac{1}{D(D+2)} \left(
\eta_{\alpha\beta}\eta_{\gamma\delta} +
\eta_{\alpha\gamma}\eta_{\beta\delta} +
\eta_{\alpha\delta}\eta_{\beta\gamma} \right) \left(l^2\right)^2 \,
\label{eq:fourlintreplace}
\end{align}
(see for example page~477 of reference~\cite{We1995}).
We find~\cite{Ve2000} that the term in the numerator of the integral
containing $k_{\mu}q_{\nu}$ may be written as
\begin{align}
k_{\mu}q_{\nu} & \bigg[ 3 M_Z^2 xy \left(2 - 3x - 6y + 4xy + 4y^2
  \right) +  3 \kkm^2 y \left( 1-x-y \right) \left( 1 - 3x - 4y + 4xy +
  4y^2 \right) +  \nonumber\\ 
 &\phantom{\bigg[} +m_f^2 \left( 5 - 3x - 20y + 20xy + 20y^2 \right) +
  \nonumber\\
 &\phantom{\bigg[} \left. + 3 l^2 \left( \left( -1 + x + 4y - 4xy - 4y^2 \right) +
  \frac{2}{D} \left( -1 + 3x + 8y - 8xy - 8y^2 \right) \right) \right] .
\end{align}
We note that the integral appears to have a divergent term. However,
it will turn out that the coefficient of the divergent term is zero
after integration over the Feynman parameters.

It is possible, although not entirely straightforward, to evaluate the
integral with the numerator given above. In order to do so, we adopt the
following strategy:
\begin{itemize}
\item Perform the integrations over the loop momentum~$l$. For the
  numerator terms not containing factors of~$l$, we use the standard
  result. For the numerator terms containing factors of~$l$, we use
  the results given in Appendix~\ref{sec:dimreginids}. We obtain a
  result that appears to diverge as $D\to 4$.
\item Order the Feynman parameter integration so as to perform the
  $x$~integral first.
\item Perform the $x$~and $y$ integration for the coefficients of the
  terms that appear to be $D \to 4$ divergent. (These are also the
  coefficients of $\gamma$ and $\log (4\pi)$ ``$\overline{\rm
  MS}$-like'' terms, which we therefore consider at the same time.)
  The terms vanish.
\item To simplify the algebra, rewrite the numerator of the $1/[m_f^2
  - y(1-x-y)\kkm^2 - xyM_Z^2]$ term to eliminate terms with
  numerator coefficients involving $m_f^2$. (Note that doing this before
  approximating the integral in the way done below makes no difference
  when making the light fermion approximation given, and is in fact
  equivalent to including an extra term in the series expansion for
  the fraction when making the heavy fermion approximation given.)
\item At this stage, there are terms in the integrand that do not
  contain masses (these arise from the step above, and from the
  ``anomalous'' terms arising from the loop integration that looks
  like equation~\eqref{eq:dimregidlsq}). Integrate these terms with
  respect to the Feynman parameters.
\item Integrate the logarithmic term with respect to~$x$ by parts,
  obtaining a contribution to the $1/[m_f^2 - y(1-x-y)\kkm^2 -
  xyM_Z^2]$ term and a logarithmic term to be integrated with respect
  to~$y$ only.
\item Integrate the remaining logarithmic term with respect to~$y$ by
  parts. One of the resultant terms is zero. Approximate the
  denominator of the other term by considering the cases $m_f^2 \ll
  M_Z^2/4$ (in which case take $m_f^2 = 0$) and $m_f^2 \gg M_Z^2/4$
  (in which case take $M_Z^2 = 0$).
\item In order to perform the remaining integral (the integrand has a
  denominator of
  \mbox{$[m_f^2 - y(1-x-y)\kkm^2 - xy M_Z^2]$}), approximate the
  denominator of the integrand by considering the cases  $m_f^2 \ll
  M_Z^2/4$ (in which case take $m_f^2 = 0$) and $m_f^2 \gg M_Z^2/2$
  (in which case take $M_Z^2 = 0$, $\kkm^2 = 0$). The latter
  approximation leads to a relatively straightforward integral. For
  the former approximation, the integral may be evaluated by
  considering the numerator and denominator as polynomials in~$x$, and
  writing the integrand in quotient+remainder form.
\end{itemize}
Following this strategy, we obtain that for each fermion, the
contribution to the amplitude coefficient $F^{(\phi)}$ is
\begin{equation}
F^{(\phi f)} = \begin{cases}
0 & \text{for $m_f^2 \ll \frac{1}{4}M_Z^2$} \, ,\\
- \frac{\omega X_f}{(4\pi)^2} \left[ \frac{8}{3} + \frac{1}{45}
  \frac{7 \kkm^2 + 11M_Z^2}{m_f^2} \right] & \text{for $m_f^2 \gg
  \frac{1}{2}M_Z^2$} \, .
\end{cases}
\label{eq:fermionintegralresult}\end{equation}

\subsubsection{W~loop diagrams}
There are no Levi-Civita terms in the individual diagram
contributions, so the overall contribution to $F_1^{(\phi)}$ is
zero. We need therefore only consider the $k_{\mu}q_{\nu}$-like terms
to obtain a contribution to the coefficient $F^{(\phi)}$.

The contributions from diagrams (f) and (j) are zero, since the
contributions only depend upon $k$, and therefore do not yield a term
of the form we are considering.

With the exception of diagrams (a), (b) and (g), the other diagrams
trivially give a zero contribution. We need to consider the
contributions from each of diagrams (a), (b) and (g).

We begin by considering the contributions from diagrams (a) and
(b). These diagrams yield an identical
contribution. We may see this through consideration of the integral
that contributes to $\diagcontpw{b}$, given in
equation~\eqref{eq:diagcontpwb}. If we take this integral, change the
integration parameter taking $l \to -l-q$, relabel the dummy suffices
and note that
\begin{align}
D_W^{\alpha\beta}(l) &= D_W^{\alpha\beta}(-l)\\
\intertext{and}
N_{\alpha\beta\gamma}(k_1,k_2,k_3) &=
-N_{\beta\alpha\gamma}(k_2,k_1,k_3) \, ,
\end{align}
we are able to obtain the expression for $\diagcontpw{a}$. We
can therefore obtain the contribution from both diagrams by
considering the form of the contribution from one of the
diagrams, and we choose diagram (b). We write the contribution in the
form
\begin{equation}
\diagcontpw{a} + \diagcontpw{b} = \frac{4i\omega}{M_W^2} \lint
\frac{S_{\mu\nu}(l)}{\left[(l-k)^2-M_W^2\right]
  \left[(l+q)^2-M_W^2\right] \left[l^2-M_W^2\right]} \, ,\\
\end{equation}
where
\begin{align}
S_{\mu\nu}(l) &= M_W^4 \eta_{\gamma\delta}
N_{\alpha\beta\mu}(l-k,-l-q,p) N_{\sigma\tau\nu}(l,-l+k,-k) \times
\nonumber\\
& \phantom{=} \times  \left( -\eta^{\alpha\tau} +
\frac{(l-k)^{\alpha}(l-k)^{\tau}}{M_W^2} \right) \left(
-\eta^{\delta\sigma} + \frac{l^{\delta}l^{\sigma}}{M_W^2} \right)
\left( -\eta^{\beta\gamma} +
\frac{(l+q)^{\beta}(l+q)^{\gamma}}{M_W^2}\right) .
\end{align}
Applying a Feynman parameterization and continuing to $D$~dimensions,
the equation becomes
\begin{align}
\diagcontpw{a} + \diagcontpw{b} &= \frac{8i\omega}{M_W^2} \md \times
\nonumber\\
&\phantom{=} \times \ldint
\int_0^1 dx \int_0^{1-x} dy \frac{S_{\mu\nu}(l+xk-yq)}{\left[ l^2 -
    M_W^2 + y(1-x-y)\kkm^2 + xyM_Z^2\right]^3} \, .
\end{align}
We leave the equation in this form for now, and turn to the
contribution from diagram (g). Unlike the corresponding diagram for
spin-2 particle decay, this diagram provides a contribution to the
amplitude coefficient. We write this contribution in the form
\begin{align}
\diagcontpw{g} &= \frac{2i\omega}{M_W^2} \frac{U_{\mu\nu}(l)}{\left[
    l^2-M_W^2 \right] \left[ (l+q)^2 - M_W^2 \right]} \, , \\
\intertext{where}
U_{\mu\nu}(l) &= M_W^4 \eta_{\sigma\tau}R_{\alpha\beta\mu\nu}
    \left(-\eta^{\tau\alpha} + \frac{l^{\tau}l^{\alpha}}{M_W^2}\right)
    \left(-\eta^{\beta\sigma} +
    \frac{(l+q)^{\beta}(l+q)^{\sigma}}{M_W^2}\right) .
\end{align}
We wish to combine this contribution with that for diagrams (a) and
(b), so we write it in the form
\begin{equation}
\diagcontpw{g} = \frac{2i\omega}{M_W^2} \lint \frac{U_{\mu\nu}(l)
  \left[(l-k)^2-M_W^2\right]}{\left[l^2-M_W^2\right]
  \left[(l+q)^2-M_W^2\right] \left[(l-k)^2-M_W^2\right]} \, .
\end{equation}
Applying a Feynman parameterization and continuing to $D$~dimensions,
the equation becomes
\begin{align}
\diagcontpw{g} &= \frac{4i\omega}{M_W^2} \md \times \nonumber\\
& \phantom{=} \times \ldint
\int_0^1 dx \int_0^{1-x} dy \frac{U_{\mu\nu}(l+xk-yq)
  \left[\left(l+(x-1)k-yq\right)^2-M_W^2\right]}{\left[ l^2 -
    M_W^2 + y(1-x-y)\kkm^2 + xyM_Z^2\right]^3} \, .
\end{align}
We may now combine the contributions from diagrams (a), (b) and
(g). Using the symmetry properties of the $l$~integral given in
equations~\eqref{eq:twolintreplace} and~\eqref{eq:fourlintreplace},
and taking a factor of $4i\omega/M_W^2$ outside the integral, we
find~\cite{Ve2000} that the term in the numerator of the integrand
containing $k_{\mu}q_{\nu}$ may be written as
\begin{align}
k_{\mu}q_{\nu}\Bigg[ & M_Z^4 \left( 2xy^2 - 2xy^3 - \frac{1}{2}x^2y -
  x^2y^2 + \frac{1}{2}x^3y - x^3y^2 \right) + \nonumber\\
 & + \kkm^2 M_Z^2 \left( 2y^2
  - 4y^3 + 2y^4 - 3xy^2 + 3xy^3 + x^2y -
  \phantom{\frac{3}{2}}\right.\nonumber\\
 & \phantom{+\kkm^2 M_Z^2 \bigg(}\left. -\frac{3}{2}x^2y^2 + 2x^2y^3 -
  x^3y + 2x^3y^2 \right) + \nonumber\\
 & + \kkm^4 \left( -xy^2 + 2xy^3 - xy^4 - \frac{1}{2}x^2y +
  \frac{5}{2}x^2y^2 - 2x^2y^3 + \frac{1}{2}x^3y - x^3y^2 \right) +
  \nonumber\\
 & + M_W^2 M_Z^2 \left( -2 + 4y - 6y^2 - 2x + 2xy + \frac{1}{2}x^2 -
  5x^2y \right) + \nonumber\\
 & + M_W^2 \kkm^2 \left( -5xy + 5xy^2 - \frac{1}{2}x^2 +
  5x^2y \right) + \nonumber\\
 & + M_W^4 \left( 12 + 8y - 8y^2 - 4x - 8xy + D\left( -8y + 8y^2 + 8xy
  \right) \right) + \nonumber\\
 & + l^2 \Bigg\{ M_Z^2 \left( \left( -2 +2y^2 + 2x + xy -
  \frac{1}{2}x^2 + 2x^2y \right) \right. + \nonumber\\
 & \phantom{l^2 \Bigg\{ M_Z^2 \Bigg( } + \left. \frac{2}{D} \left( 2 -
  4y + 4y^2 +
  x + xy - x^2 + 4x^2y \right) \right) + \nonumber\\
 & \phantom{l^2 \Bigg\{ } + \kkm^2 \left( \left( 2xy - 2xy^2 +
  \frac{1}{2}x^2 - 2x^2y \right) \right. + \nonumber\\
 & \phantom{l^2 \Bigg\{ \kkm^2 \Bigg( } \left. + \frac{2}{D} \left( -
  y + y^2 - x + 5xy - 4xy^2 + x^2 - 4x^2y \right) \right) +
  \nonumber\\
 & \left. \phantom{l^2 \Bigg\{ } + M_W^2 \left( \left( 4 + 5x \right) +
  \frac{2}{D} \left( -12 + 5x \right) \right) \right\} \nonumber\\
 & + \left( l^2 \right)^2 \left\{ -x + \frac{2}{D} \left( 1 - 2x
  \right) \right\} \Bigg] \, .
\end{align}
We evaluate this integral using a similar strategy to that for the
fermion loop case. To simplify the algebra, we rewrite the numerators
of some fractional terms in order to obtain fewer terms requiring the
application of approximations. Subdominant terms will be affected by
the order in which rewriting a
numerator and approximating the denominator is performed. However,
this effect will be negligible at the order of the approximations made
to the fractions themselves. Nevertheless, we record the order in
which we proceed through the algebra, so that it is possible to
reproduce the result.

The strategy adopted is as follows:
\begin{itemize}
\item Perform the integrations over the loop momentum~$l$. As in the fermion
case, for the numerator terms not containing factors of $l$, we use
the standard result, and for the numerator terms containing factors of
$l$, we use the results given in Appendix~\ref{sec:dimreginids}. Again
we obtain a result that, at first, appears to diverge as $D \to 4$.
\item Order the Feynman parameter integration so as to perform the
  $x$~integral first.
\item Perform the~$x$ and~$y$ integration for the coefficients of the
  terms that appear to be $D \to 4$ divergent (and of the terms that
  are ``$\overline{MS}$-like''). The terms vanish.
\item Rewrite the numerator of the
  $1/\left[M_W^2-y(1-x-y)\kkm^2-xyM_Z^2\right]$ term to eliminate
  terms with numerator coefficients involving $M_W^2$ (as in the
  fermionic case, when taken with the approximation for the
  denominator below this is equivalent to adding higher order
  corrections in the series expansion for the fraction).
\item Integrate the terms that are non-fractional and do not contain a
  logarithm.
\item Integrate the logarithmic term with respect to~$x$ by parts,
  obtaining a contribution to the
  $1/\left[M_W^2-y(1-x-y)\kkm^2-xyM_Z^2\right]$ term and a logarithmic
  term to be integrated with respect to $y$ only. Rewrite the
  numerator of the fraction to eliminate terms with coefficients
  involving $M_W^2$.
\item Integrate the remaining logarithmic term with respect to $y$ by
  parts. The only term that results is an integral with fractional
  integrand. This term can be integrated if we approximate the
  denominator by $M_W^2$. (Note that the result below does not involve
  rewriting the numerator term with $M_W^2$ coefficient for this term
  only.)
\item Perform the remaining double integral by approximating the
  denominator of the fractional integrand by $M_W^2$.
\end{itemize}
Following this strategy, we find that the contribution from the
W~boson loop diagrams to the coefficient $F^{(\phi)}$ is
\begin{equation}
F^{(\phi W)} =\kappa e g \cos \theta_W \frac{\omega}{(4\pi)^2} \left[ 14 +
  \frac{11}{15}\frac{\kkm^2}{M_W^2} - \frac{37}{15}\frac{M_Z^2}{M_W^2}
  + \frac{19}{210}\frac{\kkm^4}{M_W^4} + \frac{1}{210}\frac{\kkm^2
  M_Z^2}{M_W^4} - \frac{3}{20}\frac{M_Z^4}{M_W^4} \right] .
\label{eq:wintegralresult}\end{equation}

\subsection{Overall contribution to the spin-0 amplitude coefficients}

We have determined so far that the coefficient~$F_1^{(\phi)}$ is
zero. We have also determined that the coefficient~$F^{(\phi)}$ is
given by
\begin{align}
F^{(\phi)} &= \kappa e g \cos \theta_W \frac{\omega}{(4\pi)^2} \left[ 14 +
  \frac{11}{15} \frac{\kkm^2}{M_W^2} -
  \frac{37}{15}\frac{M_Z^2}{M_W^2} +
  \frac{19}{210}\frac{\kkm^4}{M_W^4} + \frac{1}{210} \frac{\kkm^2
  M_Z^2}{M_W^4} - \frac{3}{20}\frac{M_Z^4}{M_W^4}\right] -\nonumber\\
  &\phantom{=} -
  \frac{\kappa e Q_t g}{2 \cos \theta_W} \frac{\omega X_t}{(4\pi)^2}
  \left[ \frac{8}{3} +
  \frac{1}{45} \frac{7\kkm^2 + 11 M_Z^2}{m_t^2} \right] .
\end{align}
Substituting a value for $Q_tX_t$ (and including the colour degrees of
freedom), this becomes
\begin{align}
F^{(\phi)} &= \kappa e g \cos \theta_W \frac{\omega}{(4\pi)^2} \left[ 14 +
  \frac{11}{15} \frac{\kkm^2}{M_W^2} -
  \frac{37}{15}\frac{M_Z^2}{M_W^2}+
  \frac{19}{210}\frac{\kkm^4}{M_W^4} + \frac{1}{210} \frac{\kkm^2
  M_Z^2}{M_W^4} - \frac{3}{20}\frac{M_Z^4}{M_W^4}\right] - \nonumber\\
 & \phantom{=} - \frac{\kappa e g}{2
  \cos \theta_W} \frac{\omega}{(4\pi)^2} \left( 1 - \frac{8}{3}\sin^2
  \theta_W \right) \left[ \frac{8}{3} + \frac{1}{45} \frac{7 \kkm^2 +
  11 M_Z^2}{m_t^2} \right] ,
\end{align}
and using $M_W = M_Z \cos \theta_W$, $e = g\sin\theta_W$ and
$\sin^2\theta_W = 0.23$, along with a numerical value for
$M_Z/m_t$~\cite{Ya2006}, we obtain the approximation
\begin{equation}
F^{(\phi)} = \frac{\kappa e^2 \omega}{(4\pi)^2} \left[ 18 + 1.7
  \frac{\kkm^2}{M_Z^2} + 0.28 \frac{\kkm^4}{M_Z^4} \right] \label{eq:fp}.
\end{equation}

\section{Calculation of the decay width}\label{sec:decaywidth}

\subsection{Decays involving spin-2 Kaluza-Klein particles}

For the decay to a spin-2 KK excitation, the matrix element is given
by
\begin{equation}
\matel^{(h)} = \polgs \polps \polz F^{(h)} \left[ \left( k_{\lambda}q_{\nu}
  - k\cdot q \eta_{\nu\lambda}\right) \left( k_{\rho}q_{\mu} - k\cdot
  q \eta_{\mu\rho}\right) + \left( \lambda \leftrightarrow \rho \right)
  \right] .
\end{equation}
To obtain the square of the matrix element, we need the polarisation
sum formulae
\begin{align}
\sum_{\rm{pol}} \polgs \mathcal{E}^{\lambda' \rho'}(q) &= \frac{1}{2}
\left[  \polsumgel{\lambda}{\lambda'} \polsumgel{\rho}{\rho'} +
  \right.\nonumber\\
& \phantom{=\frac{1}{2}\Bigg[} +
  \polsumgel{\lambda}{\rho'} \polsumgel{\lambda'}{\rho} -\nonumber\\
& \phantom{=\frac{1}{2}\Bigg[}\left.- \frac{2}{3}
  \polsumgel{\lambda}{\rho} \polsumgel{\lambda'}{\rho'} \right] ,\\
\sum_{\rm{pol}} \polps \varepsilon^{\nu '}(k) &= -\eta^{\nu\nu'} \,
\label{eq:polsump},\\
\sum_{\rm{pol}} \polz \varepsilon_Z^{\mu' *}(p) &= -\eta^{\mu\mu'} +
\frac{p^{\mu}p^{\mu'}}{M_Z^2}\label{eq:polsumz}
\end{align}
(we note that the formula for the gravitational polarisation
sum~\cite{HaLZ1998,Ve1975} %p.268 of Veltman
differs from the massless case presented in~\cite{NiP2005}).
This gives~\cite{Ve2000}
\begin{equation}
\left| \matel^{(h)} \right|^2 = \left| F^{(h)} \right|^2 \left(
\frac{\left(M_Z^2-\kkm^2\right)^4 \left(7M_Z^2 +
  3\kkm^2\right)}{36M_Z^2} \right) .
\end{equation}
We note that as $\kkm \to 0$, we recover the expression obtained for
the case of decay to a massless graviton~\cite{NiP2005}, save that
using the massive gravitational polarisation spin sum causes a
difference by a factor of $7/2$ from using the massless gravitational
polarisation spin sum.

The decay width $\Gamma^{(h,\vec{n})}$ to a single spin-2 KK mode is
given by
\begin{align}
\Gamma^{(h,\vec{n})} &= \frac{p^*}{32\pi^2 M_Z^2} \int \left| \matel^{(h)}
\right|^2 d \Omega \, ,\\
\intertext{where}
p^* &= \frac{1}{2M_Z}\left(M_Z^2-\kkm^2\right).\\
\intertext{This gives}
\Gamma^{(h,\vec{n})} &= \frac{1}{576\pi M_Z^5}\left(
M_Z^2-\kkm^2\right)^5 \left(7M_Z^2+3\kkm^2\right) \left| F^{(h)}\right|^2 \, .
\end{align}
Substituting the expression for $F^{(h)}$ obtained in
equation~\eqref{eq:fh}, and recalling that $\alpha = e^2/4\pi$ and
$\kappa = \sqrt{8\pi G}$, we obtain
\begin{align}
\Gamma^{(h,\vec{n})} &= \frac{\alpha^2 G M_Z}{72\pi^2} \left( 1 -
\frac{\kkm^2}{M_Z^2} \right)^5 \left( 7M_Z^2 + 3\kkm^2 \right) \times
\nonumber\\
 &\phantom{=} \times \left[
  0.00088 + 0.27\left(\sum_{j=0}^{\infty} \frac{1}{(j+2)(j+3)(j+4)}\left( 1 -
\frac{\kkm^2}{M_Z^2} \right)^j \right) + \right.\nonumber\\
 &\phantom{= \times \Bigg[} \left. + 21 \left(
\sum_{i=0}^{\infty} \sum_{j=0}^{\infty}
\frac{1}{(i+2)(i+3)(i+4)(j+2)(j+3)(j+4)} \left( 1 -
\frac{\kkm^2}{M_Z^2} \right)^{i+j} \right) \right] .
\end{align}

To obtain the total decay width involving spin-2 graviton excitations,
we must sum
over the excitation levels. Following~\cite{HaLZ1998}, we make a continuum
approximation for the density of states at a given mass level. Given a common
compactification radius $R/2\pi$, the density of states over which we
should integrate is
\begin{equation}
\rho ( \kkm ) = \frac{R^n \kkm^{n-2}}{(4\pi)^{n/2} \Gamma(n/2)} \, ,
\end{equation}
that is, the derivative with respect to $\kkm^2$ of the volume of a
$n$-dimensional hypersphere of radius $r=\sqrt{\vec{n}^2}$. That is,
the total decay width is approximated by
\begin{equation}
\Gamma_{\rm{tot}}^{(h)} = \int_0^{M_Z^2} d\kkm^2 \Gamma^{(h,\vec{n})}
\rho(\kkm) \, .
\end{equation}
We change the integration variable to $\kkm^2/M_Z^2$ and expand the
resultant beta functions in terms of gamma functions, obtaining
\begin{align}
\Gamma_{\rm{tot}}^{(h)} &= \frac{\alpha^2 G M_Z^{3+n}R^n}{72\pi^2(4\pi)^{n/2}} \ \times
\nonumber\\
 &\phantom{=} \times \left[
  0.00088\left( \frac{7 \cdot 5!}{\Gamma(\frac{n}{2}+6)} + \frac{3
    \cdot \frac{n}{2} \cdot 5!}{\Gamma(\frac{n}{2}+7)} \right) +
  \phantom{ \left\{ \sum_{j=0}^{\infty}\right\}} \right.
  \nonumber\\ 
 &\phantom{= \times \Bigg[} +  0.27 \left\{ \sum_{j=0}^{\infty}
    \frac{1}{(j+2)(j+3)(j+4)}\left( \frac{7
    \cdot (j+5)!}{\Gamma(\frac{n}{2}+j+6)}+\frac{3 \cdot \frac{n}{2}
    \cdot (j+5)!}{\Gamma(\frac{n}{2} + j + 7)} \right) \right\} + \nonumber\\
 &\phantom{= \times \Bigg[} + 21 
\left\{ \sum_{i=0}^{\infty} \sum_{j=0}^{\infty}
\frac{1}{(i+2)(i+3)(i+4)(j+2)(j+3)(j+4)} \times \right. \nonumber\\
 &\left.\left.\phantom{= \times \Bigg[ +21.1 \sum_{i=0}^{\infty}
    \sum_{j=0}^{\infty}} \times  \left(\frac{7
    \cdot (i+j+5)!}{\Gamma(\frac{n}{2}+i+j+6)}+\frac{3 \cdot \frac{n}{2}
    \cdot (i+j+5)!}{\Gamma(\frac{n}{2} + i+j + 7)} \right)\right\}
  \right] \label{eq:spintwooverallwidth}.
\end{align}
We note that the width tends to zero as $n\to \infty$, showing that as
the number of dimensions increases for large $n$, the phase space
increase dominates the increase in the number of KK states to reduce
the decay width.

\subsection{Decays involving spin-0 Kaluza-Klein particles}

For the decay to a spin-0 KK excitation, the matrix element is given
by
\begin{equation}
\matel^{(\phi)} = \polps \polz F^{(\phi)} \left( k_{\mu}q_{\nu} - k\cdot q
\eta_{\mu\nu} \right) .
\end{equation}
Using equations~\eqref{eq:polsump} and~\eqref{eq:polsumz}, we
obtain~\cite{Ve2000}
\begin{equation}
\left| \matel^{(\phi)} \right|^2 = \frac{1}{6} \left( M_Z^2 - \kkm^2 \right)^2
\left| F^{(\phi)} \right|^2 \, .
\end{equation}
The decay width $\Gamma^{(\phi,\vec{n})}$ to a single spin-0 KK mode
is given by
\begin{align}
\Gamma^{(\phi,\vec{n})} &= \frac{p^*}{32\pi^2 M_Z^2} \int \left|
\matel^{(\phi)}
\right|^2 d \Omega \, ,\\
\intertext{where, as before,}
p^* &= \frac{1}{2M_Z}\left(M_Z^2-\kkm^2\right).\\
\intertext{This gives}
\Gamma^{(\phi,\vec{n})} &= \frac{1}{96\pi M_Z^3}\left(
M_Z^2-\kkm^2\right)^3 \left| F^{(\phi)}\right|^2 \, .
\end{align}
Substituting the expression for $F^{(\phi)}$ obtained in
equation~\eqref{eq:fp}, again using that $\alpha = e^2/4\pi$ and
$\kappa = \sqrt{8\pi G}$, and recalling that $\omega^2 = 2/(3(n+2))$,
we obtain
\begin{align}
\Gamma^{(\phi,\vec{n})} &= \frac{\alpha^2 G}{288 \pi^2 (n+2) M_Z^3}
\left(
M_Z^2-\kkm^2\right)^3 \times \nonumber\\
&\phantom{=} \times  \left[ 330 +
  63\frac{\kkm^2}{M_Z^2} + 13\frac{\kkm^4}{M_Z^4}
  +0.97\frac{\kkm^6}{M_Z^6} + 0.078\frac{\kkm^8}{M_Z^8} \right] .
\end{align}

To obtain the total decay width involving spin-0 graviton excitations,
we again approximate by an integral the sum over the excitation
levels. Recalling that we have $(n-1)$ spin-0 particles at each mass
level, the total decay width is approximated by
\begin{equation}
\Gamma_{\rm{tot}}^{(\phi)} = \int_0^{M_Z^2} d\kkm^2 \Gamma^{(\phi,\vec{n})}
(n-1) \rho(\kkm) \, .
\end{equation}
As in the spin-2 case, we change the integration variable to
$\kkm^2/M_Z^2$ and expand the resultant beta functions in terms of
gamma functions, obtaining
\begin{align}
\Gamma_{\rm{tot}}^{(\phi)} &= \frac{\alpha^2 G M_Z^{3+n} R^n}{48
  \pi^2 (4\pi)^{n/2}} \frac{(n-1)}{(n+2)} \times \nonumber\\
& \phantom{=}\times \left[
  \frac{330}{\Gamma(\frac{n}{2}+4)} - \frac{63 \cdot
  \frac{n}{2}}{\Gamma(\frac{n}{2}+5)} + \frac{13 \cdot \frac{n}{2}
  \cdot (\frac{n}{2}+1 )}{\Gamma(\frac{n}{2}+6)} - \frac{0.97 \cdot
  \frac{n}{2} \cdot (\frac{n}{2}+1) \cdot
  (\frac{n}{2}+2)}{\Gamma(\frac{n}{2}+7)} \right. - \nonumber\\
 & \phantom{= \times }\left. - \frac{0.078 \cdot
  \frac{n}{2} \cdot (\frac{n}{2}+1) \cdot
  (\frac{n}{2}+2)\cdot (\frac{n}{2}+3)}{\Gamma(\frac{n}{2}+8)} \right]
  \label{eq:spinzerooverallwidth}.
\end{align}
As in the spin-2 case, the decay width tends to zero as $n \to
\infty$. We note that the width is zero for $n=1$, which we should
expect as there are no spin-0 excited states in the case of one extra
dimension. We note also that whilst the width should be zero for
$n=0$, the expression above does not give zero for $n=0$. However, the
extra-dimensional spin sum identities of
equation~\eqref{eq:edspinsumids} are only valid for $n>0$.

\subsection{Overall decay width to photon and Kaluza-Klein graviton excitation}

We may now use equations~\eqref{eq:spintwooverallwidth}
and~\eqref{eq:spinzerooverallwidth} to write down an approximation for
the full decay width at one loop of the Z~boson to a photon and a
Kaluza-Klein excitation in the ADD scenario:
\begin{align}
\Gamma_{\rm{tot}} &= \frac{\alpha^2 G M_Z^{3+n}R^n}{72\pi^2(4\pi)^{n/2}} \ \times
\nonumber\\
 &\phantom{=} \times \left[
  0.00088\left( \frac{7 \cdot 5!}{\Gamma(\frac{n}{2}+6)} + \frac{3
    \cdot \frac{n}{2} \cdot 5!}{\Gamma(\frac{n}{2}+7)} \right) +
  \phantom{ \left\{ \sum_{j=0}^{\infty}\right\}} \right.
  \nonumber\\ 
 &\phantom{= \times \Bigg[} +  0.27 \left\{ \sum_{j=0}^{\infty}
    \frac{1}{(j+2)(j+3)(j+4)}\left( \frac{7
    \cdot (j+5)!}{\Gamma(\frac{n}{2}+j+6)}+\frac{3 \cdot \frac{n}{2}
    \cdot (j+5)!}{\Gamma(\frac{n}{2} + j + 7)} \right) \right\} + \nonumber\\
 &\phantom{= \times \Bigg[} + 21 
\left\{ \sum_{i=0}^{\infty} \sum_{j=0}^{\infty}
\frac{1}{(i+2)(i+3)(i+4)(j+2)(j+3)(j+4)} \times \right. \nonumber\\
 &\left.\phantom{= \times \Bigg[ +21.1 \sum_{i=0}^{\infty}
    \sum_{j=0}^{\infty}}\times  \left(\frac{7
    \cdot (i+j+5)!}{\Gamma(\frac{n}{2}+i+j+6)}+\frac{3 \cdot \frac{n}{2}
    \cdot (i+j+5)!}{\Gamma(\frac{n}{2} + i+j + 7)}
  \right)\right\} + \nonumber\\
 & \phantom{= \times \Bigg[ } + \frac{3}{2} \frac{(n-1)}{(n+2)}
 \left\{
  \frac{330}{\Gamma(\frac{n}{2}+4)} - \frac{63 \cdot
  \frac{n}{2}}{\Gamma(\frac{n}{2}+5)} + \frac{13 \cdot \frac{n}{2}
  \cdot (\frac{n}{2}+1 )}{\Gamma(\frac{n}{2}+6)}\right. - \nonumber\\
 & \phantom{=\times \Bigg[ \frac{3}{16} \frac{(n-1)}{(n+2)}} \left. - \frac{0.97 \cdot
  \frac{n}{2} \cdot (\frac{n}{2}+1) \cdot
  (\frac{n}{2}+2)}{\Gamma(\frac{n}{2}+7)} - \frac{0.078 \cdot
  \frac{n}{2} \cdot (\frac{n}{2}+1) \cdot
  (\frac{n}{2}+2) \cdot (\frac{n}{2}+3)}{\Gamma(\frac{n}{2}+8)} \right\}  \left] .
  \phantom{\sum_{j=0}^{\infty}}\right.\label{eq:finalanswer}
\end{align}
Two comments about this expression are inherited from the constituent
expressions of equations~\eqref{eq:spintwooverallwidth}
and~\eqref{eq:spinzerooverallwidth}. Firstly, the overall expression
does not reduce to the case of a single, massless graviton for $n=0$,
because the extra-dimensional spin sum formula in
equation~\eqref{eq:edspinsumids} is only valid for $n \geq
1$. Secondly, the overall width tends to zero as $n\to \infty$, so
that this process becomes less distinguishable from a Standard Model
background as the number of extra dimensions increases.

Tables~\ref{tab:hnumanswers} and~\ref{tab:phinumanswers} give
numerical values\footnote{Aside from the analytically approximated
  expressions for the $\tilde{\phi}$ widths, which are calculated by
  hand, these are evaluated using
Mathematica~\cite{Wo1999}. The infinite sums are evaluated by truncation once the
quoted level of significance has been achieved. The numerical
approximations to the integrals are evaluated
using the NIntegrate function with default options, and (for
comparison) using a number of
other numerical integration routines~\cite{Ha2004} with their default
options.} for the
contributions inside the square brackets in
equation~\eqref{eq:finalanswer}.
\TABULAR{| c | r @{.} l | r @{.} l | r @{.} l | r @{.} l |}{
\hline
 & \multicolumn{2}{|c|}{} & \multicolumn{6}{|c|}{\rule{0pt}{2.5ex}
  $\tilde{h}$ numerical approximation} \\
\cline{4-9}
$n$ & \multicolumn{2}{c}{\rule{0pt}{2.5ex} $\tilde{h}$ an.~approx.}
& \multicolumn{2}{|c|}{full} & \multicolumn{2}{|c|}{$m_f \to 0$} &
\multicolumn{2}{|c|}{W,t approx.}
 \\
\hline
2 & 0&16 & 0&16 & 0&16 & 0&16 \\
3 & 0&056 & 0&058 & 0&057 & 0&056 \\
4 & 0&019 & 0&020 & 0&020 & 0&019 \\
5 & 0&0066 & 0&0070 & 0&0069 & 0&0066 \\
6 & 0&0020 & 0&0023  & 0&0023 & 0&0022 \\
\hline
}{Numerical values of the $\tilde{h}$ contributions
inside the square brackets of equation~\eqref{eq:finalanswer} for some
different values of $n$ (`an.~approx.'~column), and the
equivalent values when the contributions are calculated by numerical
integration. The `full' column shows the results of integrating
numerically the full expression for the decay width (without making
the approximations made in the analytic case), and is the value with
which the analytic expression is compared. The extra columns show the
effect of performing numerical integration under the extra
approximations of neglecting the
masses of the fermions other than the top (`$m_f\to 0$' column) and of
assuming in addition to this that the W and top masses dominate the denominators of
their respective integrands (`W,t approx.'~column). The purpose of
including such columns is to give an indication of how much these
approximations change the final answer.\label{tab:hnumanswers}}%
\TABULAR{| c | r @{.} l | r @{.} l | r @{.} l | r @{.} l | r @{.} l |}{
\hline
 & \multicolumn{2}{|c|}{} & \multicolumn{8}{|c|}{\rule{0pt}{2.5ex}
  $\tilde{\phi}$ numerical approximation} \\
\cline{4-11}
$n$ & \multicolumn{2}{c}{\rule{0pt}{2.5ex} $\tilde{\phi}$ an.~approx.}
& \multicolumn{2}{|c|}{full} & \multicolumn{2}{|c|}{$m_f \to 0$} &
\multicolumn{2}{|c|}{W,t approx.}
& \multicolumn{2}{|c|}{Num.~change} \\
\hline
2 & 5&3 & 5&3 & 5&3 & 4&4 & 5&4 \\
3 & 3&9 & 4&0 & 3&9 & 3&3 & 4&0 \\
4 & 2&2 & 2&2 & 2&2 & 1&8 & 2&2 \\
5 & 1&0 & 1&0 & 1&0 & 0&85 & 1&1 \\
6 & 0&46 & 0&46 & 0&46 & 0&37 & 0&47 \\
\hline
}{Numerical values of the $\tilde{\phi}$ contributions
inside the square brackets of equation~\eqref{eq:finalanswer} for some
different values of $n$ (`an.~approx.'~column), and the
equivalent values when the contributions are calculated by numerical
integration. Again the `full' column gives the values with which those
calculated analytically are compared. The extra columns show the effects of neglecting the
masses of the fermions other than the top (`$m_f\to 0$' column), of
assuming in addition to this that the W and top masses dominate the denominators of
their respective integrands (`W,t approx.'~column), and of making this
assumption but changing the numerators to eliminate $M_W^2$ as done in
the analytic calculation (`Num.~change' column; as noted in the text
this is equivalent to considering higher order corrections in the series
expansions for the fractions in the integrand).\label{tab:phinumanswers}}%
It is notable that the contributions from the decays to spin-0
excitations are dominant.

In order to estimate the effects of the analytic approximations used
for the  integral $J(X,Y,Z)$ in
equation~\eqref{eq:hintjdefine}, and for the integrals performed to obtain
equations~\eqref{eq:fermionintegralresult} and~\eqref{eq:wintegralresult}, we
have evaluated contributions to the
decay widths by integrating numerically over
the Feynman parameters and the KK mass-squared $\kkm^2$, using
Mathematica. We find a deviation of approximately $5\%$ between the
analytically approximated and the numerically approximated answers for
the decays to spin-2 excitations, and a deviation of less than
$1\%$ for the decays to spin-0 excitations\footnote{The quoted
  deviations compare the values obtained from analytic approximation
  with the values obtained from evaluating the full integral
  numerically using Mathematica's NIntegrate routine.}. We
have also calculated numerical approximations subject to the
assumptions of zero mass for fermions other than the top, and
(additionally to the previous assumption) of the W
and top masses dominating denominators in their respective
integrands. This indicates that the zero fermion mass assumptions are
a very good approximation, but that assuming that the W and top masses
are much greater than half the Z mass yields a poorer, although still
acceptable,
approximation. It is also notable that the numerical approximations of
the integrals for the
decay into $\tilde{\phi}$ suggest that the accuracy of the analytic calculation of the
width is much improved by rewriting the numerators of the approximated
fractions as we have.

\section{Conclusions}

We have evaluated amplitudes for Z-photon-Kaluza Klein
graviton/gravi-scalar interaction, relevant to extra dimensional
models in which the Standard Model fields are confined to a 4-brane.

In addition, we have evaluated in the ADD scenario a reasonable approximation to
lowest order for the decay width of a
Z~boson to a photon and any Kaluza-Klein excitation of the graviton. This width gives an extra
contribution to the channel of Z~boson decay to photon plus missing
energy compared with the Standard Model. The channel can provide
stronger bounds on the
compactification radius $R/2\pi$ for small numbers $n$ of extra
dimensions.

The consideration of the decay channel involving spin-0 KK excitations
(gravi-scalars) has proved significant: for the decays considered,
these channels provide the larger contribution to the overall
width. The consideration of processes involving the gravi-scalar has not been common in
the literature.

We expect our signal to be most significant when the centre of mass
energy is equal to $M_Z$, as, for example, was the case at LEP\@.
We therefore intend to calculate bounds upon the compactification radius
from LEP data~\cite{AlSfuture}, given that the data appear to be in
accordance with Standard Model predictions. Our process would be
additional to the tree-level radiation of a spin-2 KK graviton tower
from a photon.

The amplitudes obtained for $Z\to \tilde{h} \gamma$ and $Z \to \tilde{\phi}
\gamma$ could also be used in the calculation of decays of a RS1
KK~mode into a Z~boson and a photon. In RS1, the KK~mode coupling is
enhanced by a warp factor, making resonant production at colliders and
interesting signature, worthy of study. Although $\tilde{h} \to \gamma Z$
and $\tilde{\phi} \to \gamma Z$ are loop-induced processes and are therefore
suppressed with respect to other, tree-level decays of a RS1 KK~mode,
the channel would be useful for checking the couplings $\kappa$ of the
excitations, \`a la reference~\cite{AlOPPSW2002}.

\acknowledgments

We should like to thank the
members of the Cambridge Supersymmetry Working Group for their
comments. KS would like to thank the Institute for Particle Physics
Phenomenology at Durham University for hospitality offered whilst this
paper was in preparation. BCA and JPS are funded by the United
Kingdom's Science and Technology
Facilities Council% (formerly the Particle Physics and Astronomy
%Research Council)
.

\appendix

\section{Dimensional regularization integral identities}\label{sec:dimreginids}

We write $D=4-\epsilon$, and in each case take the limit as $\epsilon
\to 0$. $\gamma$ is the Euler-Mascheroni constant. The integrals on
the left are in Minkowski space. The methodology used may be found,
for example, in Appendix~A.4 of reference~\cite{PeS1995}.
\begin{align}
\dimregid{l^2}{}{-\frac{1}{2}}\label{eq:dimregidlsq}\\
\dimregid{l^2/D}{\frac{1}{4}}{}\\
\dimregid{\left( l^2 \right)^2}{3X^2}{+\frac{1}{6}}\\
\dimregid{\left( l^2 \right)^2/D}{\frac{3X^2}{4}}{+\frac{2}{3}}\\
\dimregid{\left( l^2 \right)^2/(D(D+2))}{\frac{X^2}{8}}{+1}
\end{align}

\end{fmffile}
\end{document}